%% file: main.tex
\documentclass[twocolumn]{aastex61}
\usepackage{enumerate}
\usepackage{graphicx}
\usepackage{amssymb}
\usepackage{amsmath}
\newcommand{\masyr}{\ensuremath{\rm mas\,yr^{-1}}} 
\newcommand{\kms}{\ensuremath{\rm km\,s^{-1}}}
\newcommand{\ms}{\ensuremath{\rm m\,s^{-1}}}

\newcommand{\msyr}{\ensuremath{\rm m\,s^{-1}\,yr^{-1}}}

\newcommand{\rhk}{\ensuremath{R^{\prime}_{\rm HK}}}	% Activity index R'_HK
\newcommand{\logrhk}{\ensuremath{\log\rhk}}		% log of R'_HK
\newcommand{\shk}{\ensuremath{S_{\rm HK}}}	% Activity index S_HK

\newcommand{\teff}{\ensuremath{T_{\rm eff}}}
\newcommand{\logg}{\ensuremath{\log{g}}}

\newcommand{\feh}{[Fe/H]}

\newcommand{\msun}{\ensuremath{M_\sun}}

\newcommand{\dm}{\ensuremath{\Delta m}}
\newcommand{\ks}{\ensuremath{K_{s}}}
\newcommand{\lp}{\ensuremath{L'}}

\newcommand{\specmatch}{\texttt{SpecMatch}\ }
\newcommand{\emcee}{\texttt{Emcee}\ }

\begin{document}
\title{Discovery of a White Dwarf Companion to HD 159062}

\correspondingauthor{Lea~A.~Hirsch}
\email{lhirsch@berkeley.edu}

\author{Lea~A.~Hirsch}
\affiliation{Kavli Institute for Particle Astrophysics and Cosmology, Stanford University, Stanford, CA 94305, USA}
\affiliation{University of California, Berkeley, 510 Campbell Hall, Astronomy Department, Berkeley, CA 94720, USA}

\author{David~R.~Ciardi}
\affiliation{NASA Exoplanet Science Institute, Caltech/IPAC-NExScI, 1200 East California Boulevard, Pasadena, CA 91125, USA}

\author{Andrew~W.~Howard}
\affiliation{Department of Astronomy, California Institute of Technology, Pasadena, CA 91125, USA}

\author{Geoffrey~W.~Marcy}
\affiliation{University of California, Berkeley, 510 Campbell Hall, Astronomy Department, Berkeley, CA 94720, USA}

\author{Garreth Ruane}
\affiliation{Department of Astronomy, California Institute of Technology, Pasadena, CA 91125, USA}
\affiliation{NSF Astronomy and Astrophysics Postdoctoral Fellow}

\author{Erica Gonzalez}
\affiliation{Department of Astronomy and Astrophysics, University of California, Santa Cruz, CA 95064, USA}
\affiliation{NSF Graduate Research Fellow}

%\author{Robert de Rosa}
%\affiliation{Kavli Institute for Particle Astrophysics and Cosmology, Stanford University, Stanford, CA 94305, USA}

\author{Sarah Blunt}
\affiliation{Harvard-Smithsonian Center for Astrophysics, 60 Garden Street, Cambridge, MA 02138}
\affiliation{Department of Astronomy, California Institute of Technology, Pasadena, CA 91125, USA}
\affiliation{NSF Graduate Research Fellow}

\author{Justin~R.~Crepp}
\affiliation{Department of Physics, University of Notre Dame, 225 Nieuwland Science Hall, Notre Dame, IN 46556, USA}

\author{Benjamin~J.~Fulton}
\affiliation{NASA Exoplanet Science Institute, Caltech/IPAC-NExScI, 1200 East California Boulevard, Pasadena, CA 91125, USA}

\author{Howard Isaacson}
\affiliation{University of California, Berkeley, 510 Campbell Hall, Astronomy Department, Berkeley, CA 94720, USA}

\author{Molly Kosiarek}
\affiliation{Department of Astronomy and Astrophysics, University of California, Santa Cruz, CA 95064, USA}
\affiliation{NSF Graduate Research Fellow}

\author{Dimitri Mawet}
\affiliation{Department of Astronomy, California Institute of Technology, Pasadena, CA 91125, USA}
\affiliation{Jet Propulsion Laboratory, California Institute of Technology, Pasadena, CA 91109, USA}

%\author{Erik Petigura}
%\affiliation{Department of Astronomy, California Institute of Technology, Pasadena, CA 91125, USA}

\author{Evan Sinukoff}
\affiliation{Institute for Astronomy, University of Hawai‘i at Manoa, Honolulu, HI 96822, USA}

\author{Lauren Weiss}
\affiliation{Institut de Recherche sur les Exoplanètes, Dèpartement de Physique, Universitè de Montrèal,
C.P. 6128, Succ. Centre-ville, Montréal, QC H3C 3J7, Canada}

%%%%%%%%%%%%%%%%%%%%%%%%%%%%%%%%%%%%%%%%%%%%%%
\begin{abstract}

We report on the discovery of a white dwarf companion to the nearby late G dwarf star, HD 159062. The companion is detected in 14 years of precise radial velocity (RV) data, and in high-resolution imaging observations. RVs of HD 159062 from 2003--2018 reveal an acceleration of $-13.3\pm0.12$ \msyr, indicating that it hosts a companion with a long-period orbit. Subsequent imaging observations with the ShaneAO system on the Lick Observatory 3-meter Shane telescope, the PHARO AO system on the Palomar Observatory 5-meter telescope, and the NIRC2 AO system at the Keck II 10-meter telescope reveal a faint companion $2.7\arcsec$ from the primary star. We performed relative photometry, finding
$\Delta J = 10.09 \pm 0.38$ magnitudes, $\Delta \ks = 10.06 \pm 0.22$ magnitudes, and $\Delta \lp = 9.67\pm0.08$ magnitudes for the companion from these observations. Analysis of the radial velocities, astrometry, and photometry reveals that the combined data set can only be reconciled for the scenario where HD 159062 B is a white dwarf. A full Bayesian analysis of the RV and imaging data to obtain the cooling age, mass, and orbital parameters of the white dwarf indicates that the companion is an old $M_{B} = 0.65^{+0.12}_{-0.04} M_{\odot}$ white dwarf with an orbital period of $P = 250^{+130}_{-76}$ years, and a cooling age of $\tau = 8.2^{+0.3}_{-0.5}$ Gyr. 
\end{abstract}

%%%%%%%%%%%%%%%%%%%%%%%%%%%%%%%%%%%%%%%%%%%%%
\section{Introduction}
\label{section:intro}

There are more than two hundred known white dwarfs within 25 pc of the Sun \citep{Sion2014}, most of which are kinematically consistent with the thin disk population. Recent Gaia DR2 catalog queries report 139 white dwarfs within 20 pc \citep{Hollands2018} and 153 within 25 pc \citep{Jimenez-Esteban2018} based on strict cuts in Gaia color-magnitude space.
The DA spectral type, which is distinguished by Balmer lines in the spectra for \teff$>5000$ K, is the most common, with DA white dwarfs approximately twice as abundant as all other spectral types \citep{Giammichele2012,Sion2014,Kilic2018}.  

Approximately one quarter of the known nearby white dwarfs reside in binary systems, most with a less-evolved companion but several in double-degenerate binaries \citep{Holberg2016}. This fraction is notably significantly lower than the field binary fraction. However, \citet{Toonen2017} perform population synthesis modeling including stellar multiplicity and evolution, and determine that the known WD/MS binary population is actually in reasonably good agreement with models. They determine that as high a fraction as 10--30\% of single white dwarfs originate as binary systems in which the components merge during post-main sequence evolution. Low detection sensitivity to faint white dwarfs near bright, nearby main sequence stars also likely plays a role in the observed rate of WD/MS binaries \citep{Toonen2017}.
%but this may reflect a detection bias rather than a physical effect, since faint white dwarf companions to bright nearby stars have likely been missed in many surveys of the solar neighborhood. 

Understanding the population of nearby white dwarfs, and especially those in binary systems, is important for constraining the star formation history of the local neighborhood, as well as low-mass stellar evolution and white dwarf cooling models. Additionally, studying white dwarfs in binary systems can provide insights into the possible progenitors for Type 1a supernovae. Several nearby benchmark white dwarf-main sequence binary systems have been discovered with the combination of imaging and radial velocity data \citep[e.g.][]{Crepp2013a,Crepp2018}, and a new detection of such a binary system is reported here.

HD 159062 is a bright nearby G9 dwarf star with low metallicity (\feh$= -0.31$ dex). It is likely fairly old; using its observed Ca II H\&K emission diagnostic \rhk, and the empirical relation between \rhk, rotation, and age from \citet{Mamajek2008} results in an age estimate of $\approx 7$ Gyr. HD 159062 is in the solar neighborhood at 21.7 pc \citep{Gaia2018}, and as expected based on its age estimate, its kinematics are inconsistent with any nearby young clusters \citep{Gagne2018}. HD 159062 is consistent with an old main sequence G--K dwarf based on its \logg\ and \teff. Its properties are detailed in Table \ref{tab:properties}, most of which derive from the literature and are indicated with superscripts on each parameter value. 

Based on its [Fe/H] and [Mg/Fe] abundances as determined from high resolution spectroscopy, \citet{Fuhrmann2017} demonstrate that HD 159062 is most likely a Population II star. They argue that its age is most likely greater than the estimates from chromospheric activity or standard isochronal analysis would indicate, $\tau \geq 12$ Gyr. The authors list HD 159062 as a candidate blue straggler, indicating that rejuvenation from winds from a more evolved stellar companion may be complicating the standard age diagnostics.

\citet{Fuhrmann2017b} note that HD 159062 has an anomalously high barium abundance of [Ba/Fe] $= +0.4 \pm 0.01$ dex for a star of its type and metallicity. They argue that this may be further evidence of accretion of material onto HD 159062 via winds from an AGB companion which later become a white dwarf. Other abundance surveys of the solar neighborhood report more typical values for the Ba abundance of HD 159062; \citet{Mishenina2008} report [Ba/Fe] $= +0.15 \pm 0.1$ and \citet{Reddy2006} report [Ba/Fe] $= +0.17 \pm 0.12$.

There exist many previously studied examples of binary star systems in which the more massive component has evolved off the main sequence through the AGB phase and into a white dwarf. These systems are typically distinguished by enhanced abundances of s-process heavy elements in the spectrum of the less-evolved star, which are typically thought to be a result of contamination from winds or accretion from the evolved companion during its AGB phase. Barium stars and CH stars are two subsets of this category of binary systems.

Barium stars are typically observed to be G and K giants, in binary systems with a white dwarf. These binaries have typical orbital periods of 500 -- $10^4$ days, indicating that accretion from stellar winds (and not Roche-lobe overflow) is the dominant mechanism for mass transfer \citep{Boffin1988,Izzard2010,VanderSwaelmen2017}. 

The low-metallicity counterparts of Ba stars are CH stars, which are typically Population II stars with enhanced heavy-metal abundances, also due to accretion from an evolved binary companion. These types of stars were named for the lines arising from the CH molecule observed in their spectra. They are also formed via accretion from an evolved companion, and the observed binary period distribution for this class of stars is similar to that of Ba stars, with periods nearly always below $10^4$ days \citep{Jorissen2016}.

Recently, \citet{Escorza2019} published radial velocity observations and orbit fits for a compilation of 60 Ba and CH binary systems. Of these, 27 had well-measured orbits with periods ranging from hundreds to $>8000$ days. Other systems had incomplete orbital coverage, and so orbital periods were not estimated, but it can be inferred that these systems may have significantly longer orbital periods.

The inconsistent measurements of the metallicity of HD 159062 raise questions about whether or not the system could be an example of a mild Ba or CH binary. Either way, we report here that HD 159062 does definitively reside in a binary with an evolved companion.

In this paper, we report the discovery of the white dwarf companion to HD 159062, using 14 years of precise radial velocity data from Keck/HIRES as well as multi-epoch, multi-band imaging observations from the ShaneAO system at Lick Observatory, PHARO at Palomar Observatory and NIRC2 at Keck Observatory. By combining the RV and imaging data, we can constrain the orbit and cooling age of HD 159062 A and its white dwarf companion, HD 159062 B. We describe the observations and data reduction in \S \ref{section:observations}, and the astrometry and common proper motion of the companion in \S \ref{section:astrom}. We demonstrate that HD 159062 B is not consistent with a brown dwarf or main sequence companion based on its dynamics and color in \S \ref{section:notBD}. We detail our joint Bayesian analysis of the combined data set in \S \ref{section:MyMCMC}, and present the best orbital parameters. Finally, we discuss the implications for the system's evolution, especially in the context of typical Ba star systems in \S \ref{section:discussion}.

\begin{deluxetable}{lc}
    \centering
    \tablecolumns{2} 
    \tablewidth{0pt}
    \tablecaption{Properties of HD 159062
    \label{tab:properties}}
    \tablehead{\multicolumn{2}{c}{HD 159062\ Properties}}
    \startdata
         R.A. (J2000) & 17 30 16.4238 \\
         Dec. (J2000) & +47 24 07.922 \\
         %$B$ mag. & $7.961\pm0.016 ^{1}$ \\
         %$V$ mag. & $7.222\pm 0.010 ^{1}$ \\
         %$R$ mag. & 6.8$^{2}$ \\
         %$I$ mag. & 6.4$^{2}$ \\
         %$G$ mag. & 6.967$^{3}$ \\
         $J$ mag. & $5.804\pm0.018 ^{a}$ \\
         %$H$ mag. & $5.466\pm0.023 ^{4}$ \\
         \ks\ mag. & $5.392\pm0.021 ^{a}$ \\
         W1 mag. & $5.397\pm0.164 ^{b}$ \\
         $\pi$ (mas) & $46.123\pm0.024 ^{c}$ \\
         d (pc) & $21.68\pm0.01$ \\
         $\mu_{\alpha}$ (mas yr$^{-1}$) & $169.747\pm0.061 ^{c}$ \\
         $\mu_{\delta}$ (mas yr$^{-1}$) & $77.164\pm0.058 ^{c}$ \\
         \hline
         \teff\ (K) & $5283\pm100 ^{d}$ \\
         \logg\ (dex) & $4.4\pm0.1 ^{d}$ \\
         \feh\ (dex) & $-0.31\pm0.06 ^{d}$ \\
         Mass (M$_{\odot}$) & $0.76\pm0.03 ^{d}$ \\
         Radius (R$_{\odot}$) & $0.76\pm0.04 ^{d}$ \\
         \logrhk\ & -4.97 \\
         \shk & $0.170\pm0.002$\\
         V$\sin{i}$ (\kms) & $2.06\pm0.5 ^{e}$ \\
    \enddata
    %\tablenotetext{}{$^{1}$Tycho-2 \cite{Hog2000}}
    %\tablenotetext{}{$^{2}$USNO-B \cite{Monet2003}}
    \tablenotetext{}{$^{a}$ 2MASS  \cite{Cutri2003}}
    \tablenotetext{}{$^{b}$ WISE \cite{Wright2010}}
    \tablenotetext{}{$^{c}$ Gaia DR2 \cite{Gaia2018}}
    \tablenotetext{}{$^{d}$ \specmatch \cite{Petigura2017}}
    \tablenotetext{}{$^{e}$ SME \cite{Piskunov2017}}
\end{deluxetable}

%%%%%%%%%%%%%%%%%%%%%%%%%%%%%%%%%%%%%%%%%%%%%
\section{Observations and Data Reduction}
\label{section:observations}

\subsection{Radial Velocity Observations}
\label{subsection:RVs}
Using HIRES at the 10 meter Keck Telescope \citep{Vogt1994}, we have obtained 45 radial velocity observations of HD 159062 since 2003 as part of the California Planet Survey (CPS). Several of these measurements were taken as double exposures, and were later binned on a 1-day cadence.

For HD 159062, we collected spectra through the B5 or C2 decker, with a width of $0\farcs86$ and a spectral resolution of $R \approx 55,000$. The exposures were timed to yield a per-pixel SNR of $\approx 160$ at 550 nm. A template spectrum (without iodine) was taken through the B3 decker, with a width of $0\farcs57$ and spectral resolution of $R \approx 66,000$, and with per-pixel SNR of $\approx 220$. 

To reduce the spectra and extract RVs, we used the standard California Planet Survey pipeline \citep[e.g.][]{Howard2010,Howard2016}. In brief, starlight passes through a cell containing molecular iodine, imprinting a dense forest of absorption lines on the stellar spectrum. These lines are used as a high-fidelity wavelength calibration and PSF reference. The template spectrum is obtained without the iodine cell to use as a stellar spectral reference. This reference spectrum is deconvolved against the instrumental PSF, and the wavelength solution, instrumental PSF, and radial velocity are forward modeled for each observing epoch. Each spectrum is divided into $\sim 700$ ``chunks'', and the radial velocity is extracted individually for each. The RVs and uncertainties reflect the weighted average and scatter of the results for the ensemble of spectral chunks.    

\begin{figure}
    \centering
     \includegraphics[width=0.47\textwidth]{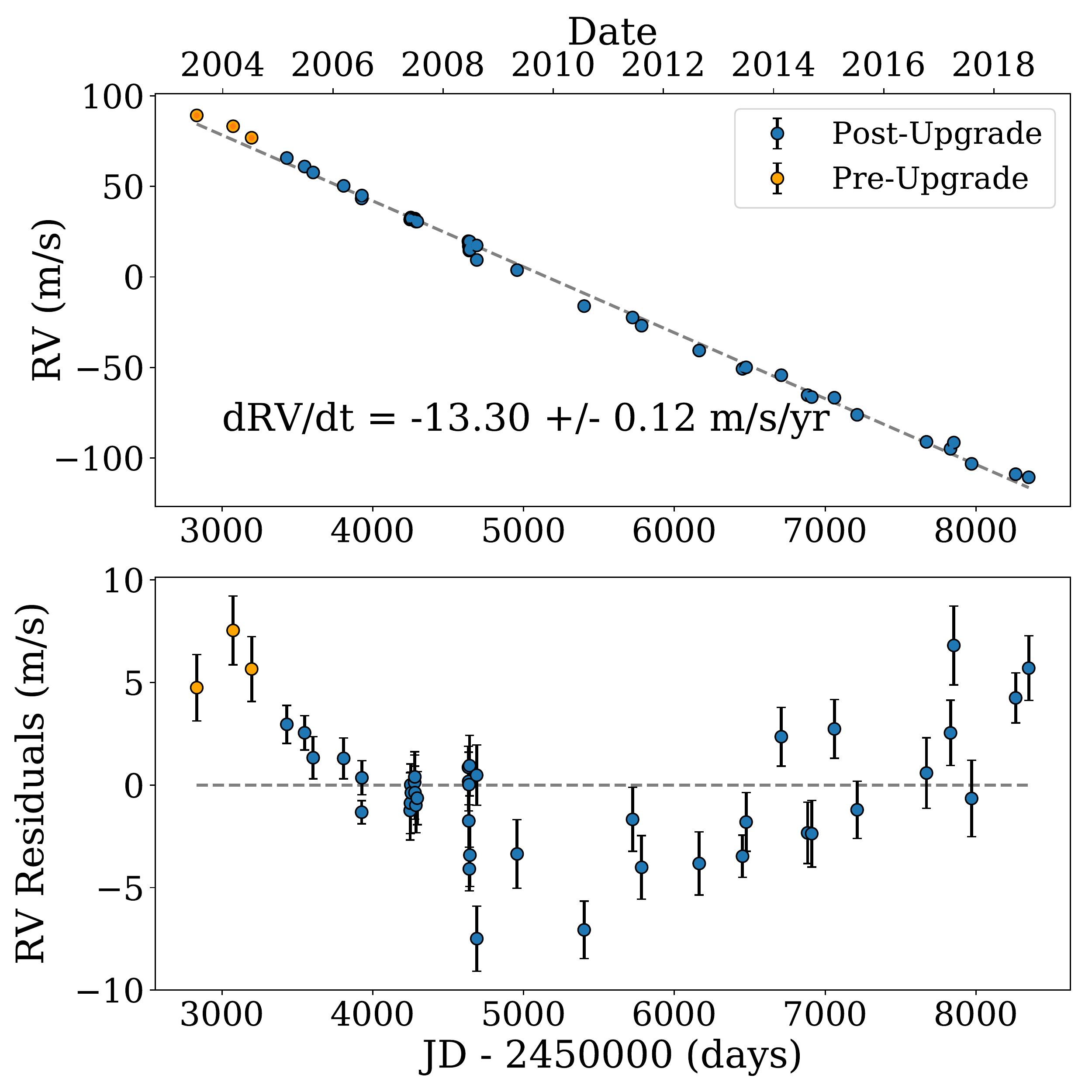}
     \caption{Top: The radial velocity time series of HD 159062 from Keck/HIRES, with a linear fit to the data. The plotted RV values have an arbitrary RV zero-point offset added, and therefore represent relative RVs, not absolute barycentric or heliocentric RVs. Data collected prior to the HIRES CCD upgrade in 2004 are plotted in orange (upper left), and are fitted with a separate RV zero-point offset in the MCMC analysis. Internal uncertainties on the RV measurements, listed in Table \ref{tab:rvs}, are smaller than the plot symbols used, so are not shown. The best-fit linear solution has a slope of $dRV/dt = -13.30\pm0.12 \ms$. Bottom: RV residuals to the best linear fit. The observations clearly show curvature over the time range of the observations.}
     \label{fig:rv_curve}
 \end{figure}

The RV time series of HD 159062 shows a strong RV acceleration of $-13.29\pm0.12$ \msyr. Careful inspection of the radial velocity time series shows slight curvature about this linear trend, which provides greater leverage on the full orbit. Figure \ref{fig:rv_curve} displays the radial velocity time series, as well as a linear fit to the data. The RV dataset is provided in Table \ref{tab:rvs}, along with the internal RV uncertainties.

 \begin{table}[]
     \footnotesize
     \centering
     \caption{Keck/HIRES Radial Velocity Data}
     \begin{tabular}{lccc}
         \hline
         {Epoch (BJD)} & {RV (\ms)} & {$\sigma$ (\ms)} & {\shk} \\
         \hline
         \input{rvs.tex}
         \hline
     \end{tabular}
     \tablenotetext{}{RV uncertainties reported here are statistical uncertainties only, calculated from the weighted variance of the RV extracted in each spectral chunk (see \S \ref{subsection:RVs}). Jitter is not included.}
     \label{tab:rvs}
 \end{table} 

\subsection{High-Resolution Imaging}

\subsubsection{Lick/ShaneAO}
We observed HD 159062 using the ShaneAO system on the 3-meter Shane Telescope at Lick Observatory \citep{Gavel2014,Srinath2014}. We obtained images in \ks\ band on UT 2016 April 20 and UT 2016 August 17. These initial images were taken as part of a uniform survey of stars in the solar neighborhood, in which the achievable contrast was improved by allowing the bright nearby primary stars to saturate the detector. With this technique, contrasts of 6--10 magnitudes were frequently obtained at separations of 1--$3\arcsec$ in under one minute of exposure time, at the expense of a saturation region including the core of the stellar PSF and occasionally the first Airy ring. Additionally, unsaturated follow-up data were required to characterize any newly discovered companions.

The saturated \ks\ images both had total integration times of 19 seconds, composed of 13 frames of 1.46 seconds each, captured in a 4-point dither pattern with a single centered setup frame included and a dither throw of 2\arcsec. The resolution at \ks\ band of the Shane Telescope is $0\farcs18$. The inner working angle of the observations, outside of the saturation region, was $\approx 0\farcs37$, and the field of view extended to $10\arcsec$ from the primary star in all directions. The plate scale was $32.6\pm0.13$ mas/pixel.

An additional image of HD 159062 was obtained on UT 2018 May 30 through the $J$ and narrow-band $CH_{4}-1.2 \mu m$ filters. The image had a total integration time of 118.3 seconds, composed of 81 1.46-second integrations captured in a 4-point dither pattern with dither throws of 2, 3, and 4\arcsec. The field of view and plate scale were the same as in the saturated \ks\ images. The typical ShaneAO correction at $J$ is significantly worse than the correction at \ks.

All images were flat-fielded and then sky-subtracted using 5 frames of dithered blank sky data of the same integration time as the science exposures. For the \ks\ images, we used a KLIP reference differential imaging algorithm to suppress the PSF of the primary star, using all \ks\ images of other stars obtained on the same night as the target exposure as a reference library \citep{jasonwang2015}. However, the companion was sufficiently separated from the primary star's PSF at $2.7\arcsec$, such that the local image statistics were not significantly affected by the PSF subtraction. 

In the initial images, we detected a faint companion at $\approx 2.7\arcsec$ from the primary star. In the first image from April 2016, the signal-to-noise ratio of the detection was low, $SNR = 3.5$. In the second image from August 2016, $SNR = 5.6$. The persistence of this low SNR companion over several months indicated its physical nature, but precise relative photometry could not be calculated from the $K$-band data due to the primary star's saturation. Therefore, we pursued follow-up imaging observations. 

In the unsaturated $J + CH_4$ image, the companion detection had $SNR = 8.4$. We performed aperture photometry to obtain the $J$ contrast, using an aperture radius of 1 FWHM $= 3.9$ pixels $= 0\farcs13$ to extract relative flux values for the primary star and faint companion. A sky annulus spanning 3-5 FWHM was used.

Because of the brightness of the primary star, we extracted the flux of the companion using a high-pass filtered version of the image, to suppress the background flux at the location of the companion due to the bright primary. 

We calibrated the companion flux measurement by injection/recovery. One hundred companions with similar contrast to the true companion were injected into our image at 2.7\arcsec from the primary and at various position angles, yielding synthetic companion images, which were then high-pass filtered with the same window size as was used for the true companion photometry. The flux ratios for the injected companions were then extracted and compared to the injected flux ratios. We determined the median multiplicative correction required, and this correction was applied to the aperture photometry for the true companion. The scatter in measured flux for the injected companions was added in quadrature to the uncertainty on our actual photometric data point.

\begin{figure*}
    \centering
     \includegraphics[width=0.7\textwidth,trim={1.9cm 0 2.5cm 0},clip]{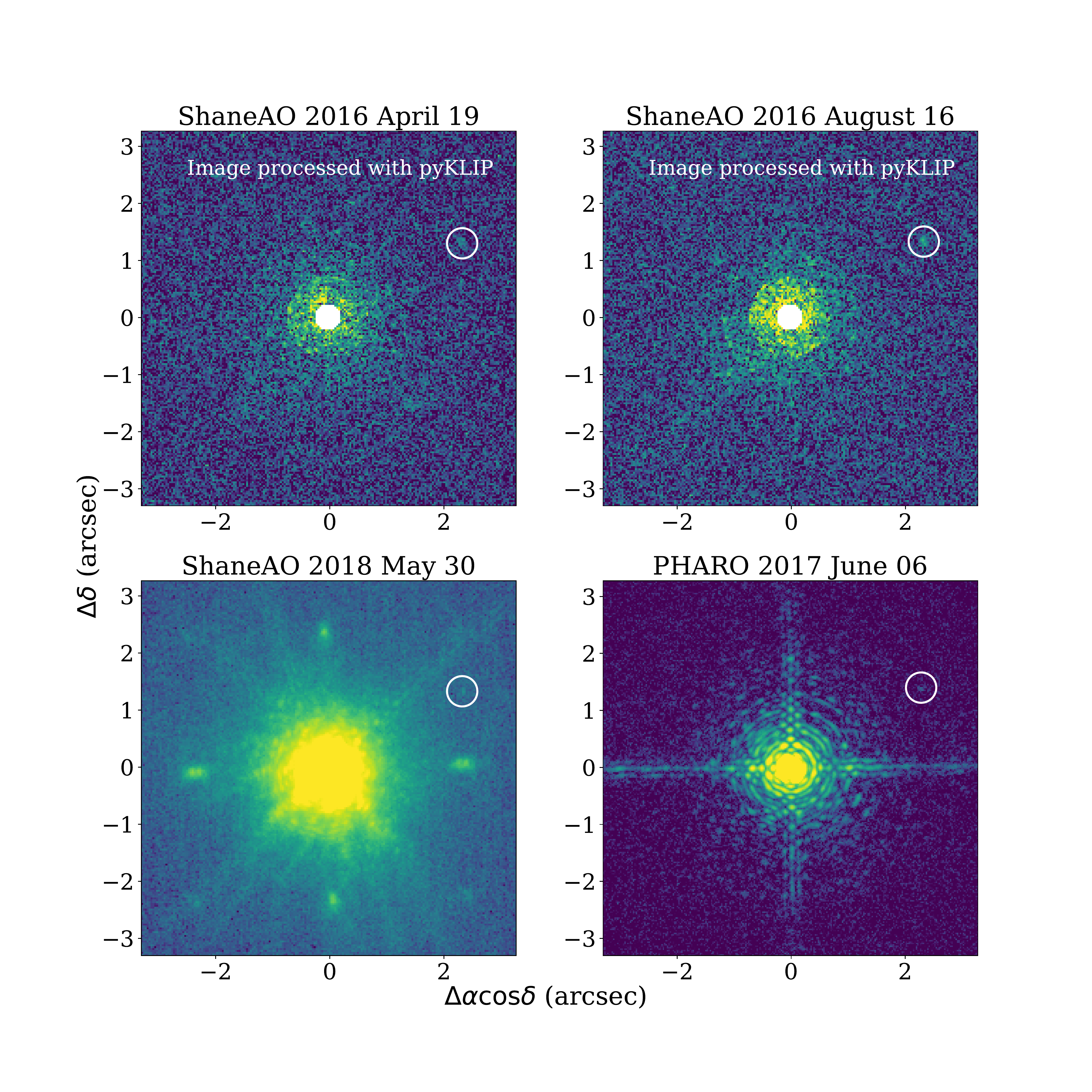}
     \caption{First four imaging epochs in which we detect the faint companion to HD 159062. The first two epochs are from ShaneAO, and the companion is detected in \ks\ band but the primary star is fully saturated, so accurate photometry cannot be computed. Each of these images was processed with pyKLIP \citep{jasonwang2015} to suppress the starlight from the central star. The third image is an unsaturated image from ShaneAO through the $J$ and $CH_{4}-1.2 \mu m$ filters, which allowed the measurement of aperture photometry yielding a $J$-band contrast of $\Delta J = 10.09\pm0.38$ mag. The fourth image is an unsaturated image from Palomar/PHARO through the \ks\ and $H_2\ 2 - 1$ filters which allowed us to measure $\Delta \ks = 10.06\pm0.22$ mag. }
     \label{fig:images}
 \end{figure*}

This methodology yielded a measured contrast of $\Delta J = 10.09\pm0.38$ magnitudes. Here, we treat the contrast measured in the narrow-band $CH_4-1.2 \mu m$ filter as a proxy for the $J$-band contrast. The ShaneAO images are shown in Figure \ref{fig:images}. The $J$-band image was not processed with the KLIP algorithm due to a lack of other $J$-band images from the night of the target observations to serve as a PSF reference library.

\subsubsection{Hale/PHARO}
We observed HD 159062 with the PHARO adaptive optics system on the 200\arcsec\ Hale Telescope at Palomar Observatory \citep{Hayward2001} on UT 2017 June 7. The observations were carried out in the \ks\ band with the narrow $H_2\ 2-1$ filter in the grism wheel. The total integration time was 21 seconds, composed of 15 frames of 1.416 seconds each, captured in a 5-point dither pattern with a dither throw of 3--5\arcsec. The pixel scale of the PHARO observations was 25.0 mas/pixel and the diffraction limit at the wavelength of observation was $0\farcs11$.

The bright primary star was unsaturated in these integrations. The frames were flat-fielded, and a sky background image was constructed from the median of the dithered images and subtracted. The companion was recovered at $SNR=6.5$ in these Palomar data, and the unsaturated core of the primary PSF allowed us to calculate the relative brightness for this companion.

Aperture photometry was calculated on the unsaturated Palomar image using the same methodology as in the ShaneAO $J$-band image, with an aperture radius of 1 FWHM $ = 3.65$ pixels $= 0\farcs09$. This resulted in a contrast of $\Delta \ks = 10.06\pm0.22$ for the companion, which is located approximately 30 FWHM away from the primary.

The imaging detection from PHARO is displayed in Figure \ref{fig:images}, with the faint companion indicated by the white circle.

\subsubsection{Keck/NIRC2}
\label{subsubsec:nirc2}
We followed up HD 159062\ in \lp\ band (3.8 $\mu$m) with Keck NIRC2 \citep{Wizinowich2000,Wizinowich2006,VanDam2006} on UT 2018 April 27 with natural guide star adaptive optics and the vortex coronagraph \citep{Serabyn2017}. The angular resolution was $0\farcs08$ and the plate scale was $9.942\pm0.05$ mas per pixel \citep{Service2016}. The field rotator was set to vertical angle mode, such that the telescope pupil tracks the elevation axis, to enable angular differential imaging \citep{Marois2006}. We took 85 frames with a discrete integration time of 0.5 seconds and 30 coadds resulting in a total integration time of 21.3 minutes over a 90-minute observation. During that time, we also took five images of the off-axis point spread function (PSF) with a discrete integration time of 0.008 seconds and 100 coadds as well as five images of the sky background with integration times matching that of the science and off-axis PSF frames. The sky off-axis PSF and background frames were taken every 15-20 min during the observing sequence. The alignment of the star and the center of the vortex focal plane mask was maintained by the QACITS tip-tilt control algorithm \citep{Huby2017}. 

Bad pixels identified in dark frames and sky flats were replaced by the median of neighboring values. The sky flat was the median of 10 images of a blank patch of sky with the coronagraph focal plane mask removed (0.75 second discrete integrations, 10 coadds each). We subtracted the sky background frames from each frame individually using a scale factor to account for background variability. The frames were centered based on the position of the optical vortex core in the median of the science frames and each of the individual frames were co-registered using the peak of the cross correlation with the median frame. 

We applied principal component analysis \citep[PCA,][]{Soummer2012} to estimate and subtract the stellar contribution (on-axis PSF) from the images using the Vortex Image Processing (VIP) software package \citep{Gomez2017}. The final data product is a cube of median science frames with the starlight subtracted using 1 to 50 principal components (PCs). The significance of the imaged companion was maximized by subtracting the model of the on-axis PSF using 8 PCs. The PCs were computed within an annulus about the star centered on the companion with a width of 4 times the full width half maximum (FWHM) of the off-axis PSF (FWHM = 8 pixels). 

To compute the photometry and astrometry of the companion, we subtracted a copy of the off-axis PSF at the location of the white dwarf in each science frame, varied its location and brightness, and repeated the PCA reduction with 8 PCs until the values were minimized in the final reduced images in a 2×FWHM radius about the companion’s position using a downhill simplex algorithm. To estimate the error, we re-injected the best-fit model of the companion into the pre-processed science frames at the same separation and brightness, but varied the parallactic angle by 10 degrees and retrieved the photometry and astrometry using the same method 36 times tracing a full circle about the star. We used the standard deviation of the measured flux and position of the injected companions as the uncertainty on each parameter. 
The NIRC2 narrow camera distortion correction was then applied by determining the location of the companion in each frame, interpolating the x-axis and y-axis distortion maps to the companion position, and then recalculating the astrometry of the primary star and companion. The scatter in the distortion map at the companion locations through the ADI sequence were added to the astrometric uncertainties in quadrature.

\begin{figure}
     \centering
     \includegraphics[trim={0 0 0 0.75cm},clip,width=0.5\textwidth]{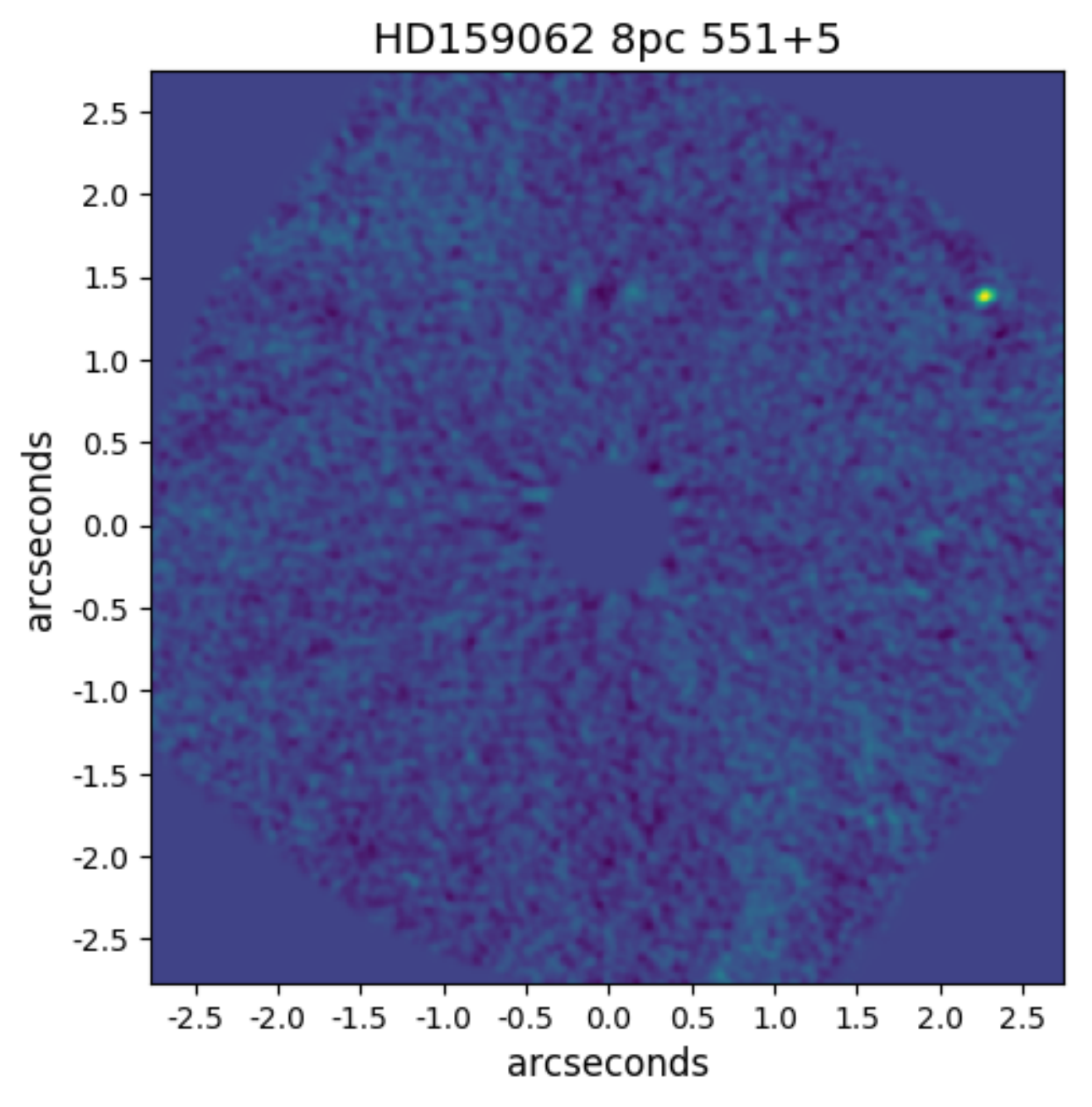}
     \caption{\lp-band image of HD 159062 B from Keck/NIRC2 using the vector vortex coronograph to suppress the starlight of HD 159062. From this image, we measure a contrast of $\Delta \lp = 9.67\pm0.08$ mag.}
     \label{fig:lband_im}
 \end{figure}

The companion was detected at an SNR of 17.5 with a flux ratio and angular separation between the companion and star of $\Delta\lp=9.65\pm0.06$ and $2\farcs673 \pm 0\farcs003$, respectively. The parallactic angle from north towards east was $301.27^{\circ} \pm 0.05^{\circ}$. The final reduced image is displayed in Figure \ref{fig:lband_im}. 

In addition to the new NIRC2 image obtained for this analysis, archival data on HD 159062 from UT 26 June 2012 and UT 12 October 2014 were discovered in the Keck Observatory archive. These data were collected as part of the TRENDS high contrast imaging survey, and both epochs included a high SNR detection of the faint companion. 

HD 159062 was observed through the TRENDS
survey \citep{Crepp2012} on 2012 June 24 UT and 2014 October 12 UT. Observations were taken with the narrow mode camera setting with plate scale $9.952 \pm 0.002$ mas
pix$^{-1}$ \citep{Yelda2010} using the full 1024” x 1024” FOV. Images were processed using standard flat-fielding, sky background subtraction, and focal plane distortion correction applied as prescribed in \citet{Yelda2010}.

The UT 2012 June 24 observations consisted of 10 frames taken with integration times of 1 second per coadd and 10 coadds, collected in position angle mode. These observations were taken under clear conditions and at an airmass of 1.1 with the $K'$ filter. The star was obscured by the 300 mas corona300 coronograph.

The UT 2014 October 12 observations consisted of 76 usable frames with integration times of 4 seconds per coadd and 6 coadds per frame. These data were taken under clear conditions and at an airmass of 1.5 with the \ks\ filter and the instrument in vertical angle mode. In this epoch the 600 mas corona600 coronograph was used to obscure the bright primary star. Prior to measuring accurate separation and position
angle, each image was de-rotated using the prescription of \citet{Yelda2010}.

Astrometric information was extracted from the archival NIRC2 images, but due to duplication of filters and lack of unocculted images, photometry was not calculated from these observations. 

  \begin{figure*}
    \centering
     \includegraphics[width=0.7\textwidth,trim={0 0 0 0},clip]{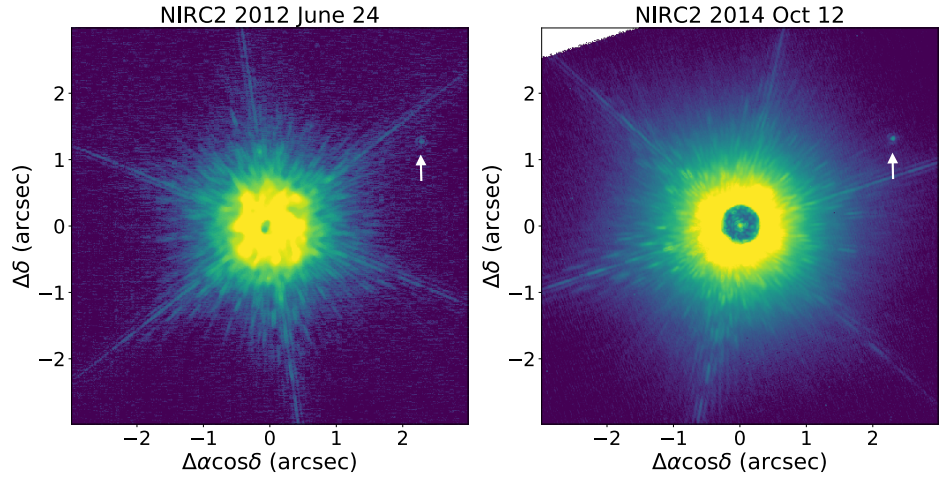}
     \caption{Archival NIRC2 images of HD 159062 and its white dwarf companion from 2012 and 2014 in the \ks\ band behind the standard coronographic spots. In 2012, the 300 mas coronograph mask was utilized, while the 600 mas coronograph was used in 2014. The companion is detected at high SNR in both epochs, and relative astrometry is reported in Table \ref{tab:imaging}.}
     \label{fig:images_archival}
 \end{figure*}
 
 \begin{table*}[t]
     \centering
     \caption{Summary of imaging observations}
     \begin{tabular}{lllccc}
        \hline
        {\bf Epoch (JD)} & {\bf Instrument} & {\bf Filter} & {\bf Separation ($\arcsec$)} & {\bf P.A. (deg)} & {\bf Contrast ($\Delta$-mag)}  \\
        \hline
        2456102.89193 & NIRC2 & $K'$ & $2.594\pm0.014$ & $298.6\pm0.1$ & \nodata \\
        2456942.70077 & NIRC2 & \ks & $2.637\pm0.014$ & $299.5\pm0.1$ & \nodata \\
        2457498.99002 & ShaneAO & \ks & $2.66\pm0.03$ & $301.5 \pm 0.56$ & \nodata \\
        2457617.69501 & ShaneAO & \ks & $2.68\pm0.02$ & $301.7 \pm 0.47$ & \nodata \\
        2457911.93061 & PHARO & \ks & $2.671\pm0.004$ & $301.3\pm0.2$ & $10.06\pm0.22$ \\
        2458236.09489 & NIRC2 & \lp & $2.673\pm0.003$ & $301.27\pm0.04$ & $9.67\pm0.08$ \\
        2458268.86233 & ShaneAO & $J$ & $2.67\pm0.01$ & $301.0\pm0.24$ & $10.09\pm0.38$ \\
     \end{tabular}
     \label{tab:imaging}
 \end{table*}

%%%%%%%%%%%%%%%%%%%%%%%%%%%%%%%%%%%%%
\section{Astrometry and Common Proper Motion of Imaged Companion}
\label{section:astrom}
%{\bf Astrometric data was extracted from each imaging observation, and is plotted and tabulated in Figure \ref{fig:proper_motion_zoom} and Table \ref{tab:imaging}. Due to inconsistencies in instrumental setup and unknown variations in plate scale and true north angle, the astrometry from different instruments and epochs is difficult to combine. In our orbit analysis, described in \S \ref{section:MyMCMC}, we therefore limit the astrometric analysis to only the NIRC2 astrometric data.}

The position of the companion in the saturated ShaneAO images, and the positions of both the primary and companion in the unsaturated  ShaneAO, PHARO, and archival NIRC2 images, were extracted using the \texttt{python} implementation of DAOStarFinder in the \texttt{photutils} package \citep{bradley2017}. 

For the PHARO data, we adopted an uncertainty on the position of the primary star of 0.1 pixel, consistent with previous astrometric efforts on this instrument (D. Ciardi, private communication). Since the companion was detected at low SNR, we use a heuristic estimate for uncertainty, $\sigma \approx \mathrm{FWHM}/2.355/\mathrm{SNR} = 0.24$ pixel. The plate scale of the Palomar AO instrument has been measured to be $25.09\pm0.04$ mas/pixel by \citet{Metchev2006}, and the rotational position of the instrument has been shown to be stable to within $0.12^{\circ}$. These uncertainties are added in quadrature to the astrometric datapoints.

For the unsaturated ShaneAO $J$-band image, we assume an astrometric precision of 0.2 pixel for both the primary and companion. The plate scale for the ShaneAO instrument is $32.6\pm0.13$ mas/pixel and the field rotation has recently been measured to be $1.87^{\circ}\pm0.13^{\circ}$ (G. Duch\^ene, private communication). Once the separation and position angle were calculated from the ShaneAO data, $1.87^{\circ}$ were added to the P.A. measurements, and the rotational and plate scale uncertainties were added in quadrature.

For the saturated ShaneAO images, the position of the saturated primary star was determined by rotational symmetry. In brief, we calculated the residuals between the reduced image and a rotated version of the image, for several different rotations about each prospective center pixel. We interpolated this residual map to find the optimal rotational centroid for the primary star. This method was adapted from \citet{Morzinski2015}. 
We adopt 0.5 pixel uncertainty for the primary star due to its saturation, and 0.4 pixel for the secondary based on the heuristic diagnostic.  
The derivation of the astrometric position of the companion in the new NIRC2 image is described in \S \ref{subsubsec:nirc2}. For the archival NIRC2 data, we adopted a 0.1 pixel uncertainty on the position of the companion, and due to the difficulty of centroiding the primary star behind the coronographic masks, we adopted a 0.5 pixel uncertainty on the position of the primary in both epochs. We added the uncertainty in the plate scale and true north angle \citep{Yelda2010} in quadrature to our separation and P.A. measurements.

\begin{figure*}
    \centering
     \includegraphics[width=0.7\textwidth,trim={0 2.8cm 0 0},clip]{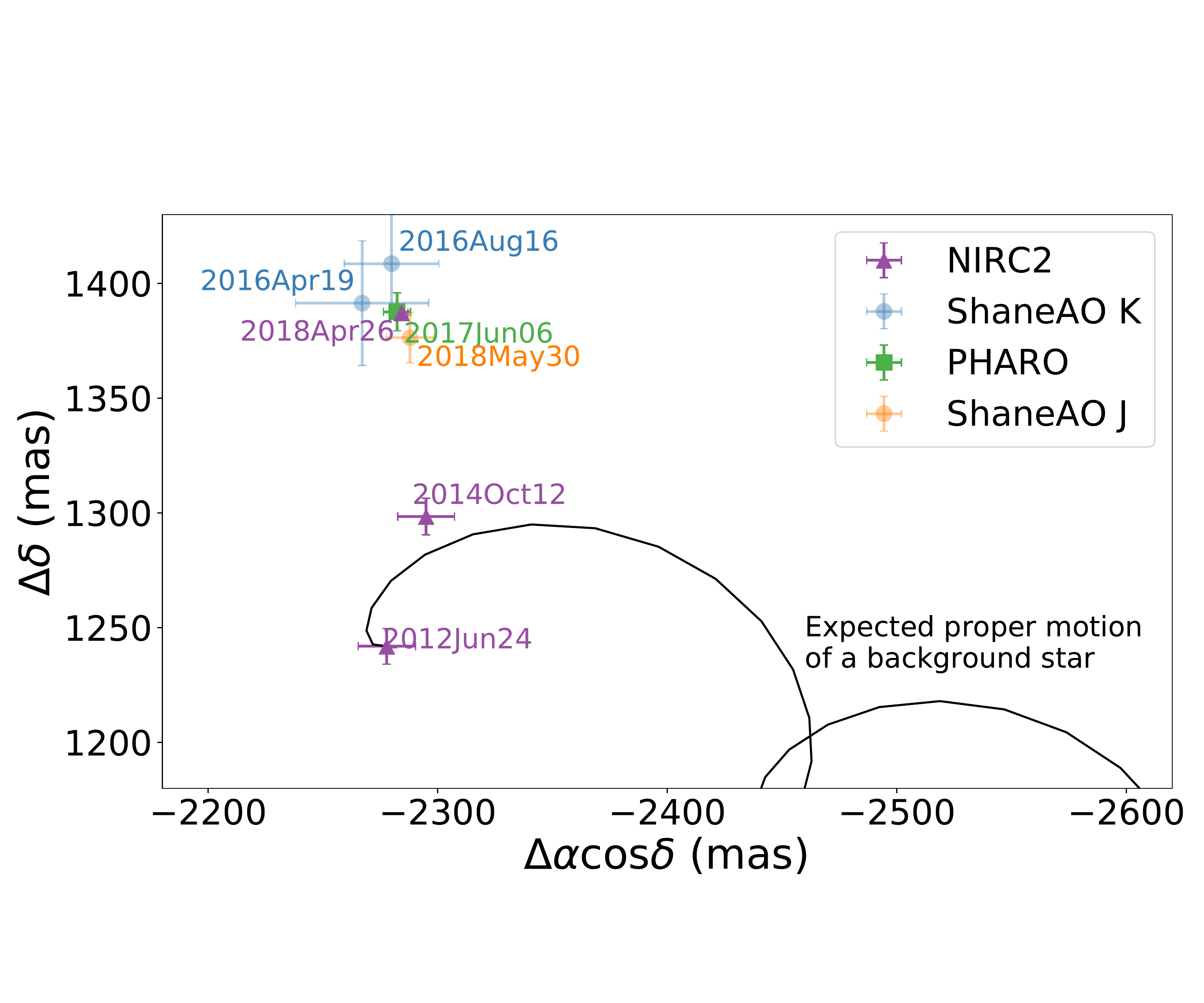}
     \caption{Astrometry calculated from the imaging observations. The relative motion of a stationary background source due to the parallax and proper motion of HD 159062 is indicated by the black curve. The position of the imaged companion does not follow this predicted track, so is not consistent with a very distant background star.}
     \label{fig:proper_motion_zoom}
 \end{figure*}

Our astrometry of the imaged companion to HD 159062 spans six years, allowing us to check for common proper motion between the primary and imaged companion. HD 159062 has a proper motion of $\Delta \mathrm{R.A.} = 169.747\pm0.061$ \masyr\ and $\Delta \mathrm{Dec} = 77.164\pm0.058$ \masyr\ \citep{Gaia2018}, large enough to allow us to detect the relative motion of a background source. Figure \ref{fig:proper_motion_zoom} shows the expected relative motion of a background source due to the proper motion of HD 159062 as well as the observed astrometry. We conclusively rule out the scenario that the imaged companion is a distant background stellar source.

Since it is likely bound, HD 159062 B must be located at the same distance from Earth as its primary star. Combining this distance with its observed photometry rules out a main sequence dwarf scenario for the companion; for reference, an M8V star at the bottom of the main sequence would have an expected contrast of only $\Delta \ks \approx 6.5$ magnitudes with the G9 primary, much less than the measured contrast of HD 159062 B. Instead, this companion must be an intrinsically faint object, either a brown dwarf or white dwarf.

%%%%%%%%%%%%%%%%%%%%%%%%%%%%%%%%
\section{HD 159062 B is not a Brown Dwarf}
\label{section:notBD}

A simple argument invoking acceleration rules out the brown dwarf scenario for HD 159062 B based on the mass constraint from the radial velocity acceleration of HD 159062 A and the projected separation between the components.

Rearranging Newton's second law and the Newtonian law of gravity, for a companion B orbiting a primary star A we can equate the magnitude of the force vector and the magnitude of the acceleration, such that:
\begin{equation}
    M_{B} = \frac{a_{A} r^{2}}{G}
    \label{eq1}
\end{equation}
where $M_{B}$ is the mass of the companion, $a_{A}$ is the instantaneous acceleration of the primary, $r$ is the physical separation between the two components, and $G$ is the gravitational constant. 

The lower limit for the instantaneous acceleration of the primary star comes from the slope of its radial velocities, $ dRV/dt  = -13.29 \pm 0.12$ \msyr. Since this describes only the acceleration in the radial direction, and does not account for acceleration along the plane of the sky, $\left | dRV/dt \right |$ is a lower limit for the full acceleration, or $a_A \geq \left | dRV/dt \right |$. 

Likewise, the lower limit for physical separation is the projected separation of the two components, $\rho_{\rm proj} = 57.78 \pm 0.07$ AU. This is also a projection of the full separation, so $r \geq \rho_{\rm proj}$. 

By substitution into Equation \ref{eq1}, we find:

\begin{equation}
    M_{B} \geq \left |\frac{dRV}{dt} \right | \rho_{proj}^{2} G^{-1} \geq 0.24 \msun.
\end{equation}

This mass, derived from a simple acceleration argument, represents a dynamical lower limit for the companion, placing it securely above the maximum mass for a brown dwarf. 

HD 159062 B's infrared colors confirm that it is not a brown dwarf. Since \lp\ photometry was not readily available for HD 159062 A, we use the WISE W1 measurement of HD 159062 A as a proxy for \lp. This is reasonable since the $W1-\lp$ color has been shown to be zero for stars earlier than M0 \citep{DeRosa2016}. From the apparent magnitudes of the primary star and the measured contrasts, we find that HD 159062 B has apparent magnitudes of $J = 15.89\pm0.38$ mag., $\ks = 15.45\pm0.22$ mag., and $\lp = 15.07\pm0.18$ mag. Its colors are therefore $J-\ks = 0.44\pm0.44$ mag. and $\ks-\lp = 0.38\pm0.28$ mag.

Low-mass stars and brown dwarfs spanning spectral types from M0 through T6 have $\ks-\lp$ colors ranging from 0.5 to 2.0 magnitudes, marginally consistent with the measured $\ks-\lp$ color of HD 159062 B. However, these objects also have $J-\ks$ colors ranging from 0.8 to 1.8 magnitudes \citep{Leggett2001}, significantly redder than the observed $J-\ks$ color of HD 159062 B. 

We conclude that HD 159062 B is a white dwarf. Both the mass lower limit and the infrared color of the faint companion are consistent with this scenario. HD 159062 B is too massive to be a brown dwarf, and too faint to be a main sequence dwarf. Additionally, the predictions from \citet{Fuhrmann2017b} regarding the enhanced Ba abundance of HD 159062 A, outlined in \S \ref{section:intro}, strengthen the white dwarf interpretation.

%%%%%%%%%%%%%%%%%%%%%%%%%%%%%%%%%%%%%%
\section{Orbital Analysis}
\label{section:MyMCMC}

We performed a joint MCMC analysis of the 45 radial velocity data points and a subset of the astrometric data. Since the imaging observations were taken on three different instruments, uncertainties and variations in the field rotation and plate scale of the instruments make combining the astrometric data difficult. We therefore chose to limit our analysis to the three NIRC2 data points, which we expected to be the most consistent in plate scale and rotation, and which conveniently covered the largest time baseline of the orbit.

The likelihood function we used is:
\begin{multline}
    \log{\mathcal{L}} = -\sum_i{\left[\frac{(v_i-v_m(t_i))^2}{2(\sigma_i^2 + \sigma_{\rm jit}^2)} + \log{\sqrt{2\pi(\sigma_i^2+\sigma_{\rm jit}^2)}}\right]}\\
    -\sum_j{\left[\frac{(x_j-x_m(t_j))^2}{2\sigma_j^2} + \log{\sqrt{2\pi\sigma_j^2}}\right]}
\end{multline}

Here, $v_i$ ($t_i$) and $x_j$ ($t_j$) correspond to the radial velocity and astrometric measurements (epochs), where a zero-point offset $\gamma$ has been added to the RV data as a free parameter in the model. Due to a HIRES CCD upgrade in 2004, we use one $\gamma$ parameter for the first three RV data points which were taken prior to the upgrade ($\gamma_k$), and another for the post-upgrade data ($\gamma_j$). $v_m(t_i)$ and $x_m(t_j)$ are the model velocities and positions at the RV and astrometric epochs respectively. $\sigma_i$ is the internal RV precision, and $\sigma_{\rm jit}$ is the additional uncertainty added in quadrature due to stellar jitter and instrumental uncertainties not included in the internal precision estimated from the RV chunks. $\sigma_j$ is the astrometric uncertainty.

We implemented some standard priors on the sample parameters for this fit, detailed in Table \ref{tab:priors}. We sampled from a uniform mass distribution between our dynamical lower limit, $0.24\msun$ from \S \ref{section:notBD}, and the Chandrasekhar upper limit for white dwarf mass, $1.44 \msun$. We chose to sample in $\log{P}$ as a variant of a Jeffreys prior for the orbital period since it is a scale parameter. We chose our lower bound on the orbital period to be several times lower than the minimum orbital period calculated by assuming the semi-major axis is equivalent to the measured projected separation. The projected physical separation we measured was approximately 60 AU, which would correspond to a period of  $\approx 400$ years assuming a total system mass of 1.3\msun. The upper bound on period was conservatively chosen to be $10^{4}$ years.

\begin{table}[h]
    \centering
    \caption{MCMC Priors on Orbital Parameters}
    \begin{tabular}{ll}
        \hline
        {\bf Parameter} & {\bf Prior} \\
        \hline
        m$_{A}$ & Gaussian, $0.80 \pm 0.05\ M_{\odot}$ \\
        $\mathrm{m}_{B}$ & Uniform, $[0.24, 1.44]\ M_{\odot}$ \\
        $\log{\mathrm{P}}$ & Uniform, $[10^2, 10^4]$ yr \\
        $\sqrt{e}\cos{\omega}$ & Uniform, $[0.0, 1.0]$ \\
        $\sqrt{e}\sin{\omega}$ & Uniform, $[0.0, 1.0]$ \\
        %e & $< 1.0$ \\
        $\Omega$ & Uniform, $[0, 2\pi]$ \\
        $M_{0}(t_0)$ & Uniform, $[0, 2\pi]$ \\
        $\cos{i}$ & Uniform, $[-1.0, 1.0]$ \\
        $\pi$ & Gaussian, $46.123\pm0.024$ mas \\
        $\gamma_j - \gamma_k$ & Gaussian, $0\pm2\ \ms$
        
    \end{tabular}
    \label{tab:priors}
\end{table}

We sampled in $\sqrt{e}\cos{\omega}$ and $\sqrt{e}\sin{\omega}$ rather than in eccentricity and argument of periastron to improve convergence time while maintaining a uniform prior on the eccentricity. Our priors on longitude of the ascending node $\Omega$ and mean anomaly $M_{0}$ at the epoch of the first observation were uniform between 0 and $2\pi$, as we have found that this improves convergence time over allowing these parameters to be unconstrained and later taking the modulus of $2\pi$ (Robert DeRosa, private communication). We sampled uniformly in $\cos{i}$ for uniform distribution of inclination over a sphere. 

Previous studies have made use of RV observations of standard stars to place the HIRES pre-upgrade and post-upgrade data on a uniform scale. We therefore place a Gaussian prior on the difference between the pre-upgrade velocity zero point $\gamma_k$ and the post-upgrade zero point $\gamma_j$ such that $\gamma_j - \gamma_k = 0\pm2\ \ms$.

The priors on primary mass m$_{A}$ and parallax $\pi$ were Gaussian. The parallax constraint is from the Gaia DR2 astrometric results \citep{Gaia2018}. For the primary mass, we based our prior on both a new spectroscopic estimate of the mass of HD 159062 A and on literature values. We use the stellar spectroscopic analysis tool \specmatch \citep{Petigura2017}, which performs fits to theoretical stellar spectra to obtain stellar parameters \teff, \logg, and \feh. These parameters are then converted to stellar mass and radius using the package \texttt{isoclassify} \citep{Huber2017}. This results in a mass estimate for HD 159062 A of $0.76\pm0.03$ \msun. 

We also perform a literature search for measurements of the mass of HD 159062 A. \citet{Ramirez2012,Ramirez2013} report a mass of m$_{A} = 0.8^{+0.02}_{-0.01} \msun$; \citet{Brewer2016} report masses of m$_{A} = 1.01\pm0.14 \msun$ or m$_{A} = 0.8\pm 0.02 \msun$ depending on which mass derivation method is used; \citet{Casagrande2011} reports m$_{A} = 0.84^{+0.04}_{-0.02} \msun$; and \citet{Fuhrmann2017b} reports m$_{A} =  0.84$ \msun. All literature mass measurements agree within $3\sigma$ with our spectroscopic measurement, but all indicate slightly more massive solutions. Therefore, we adopt a prior for HD 159062 A of $\mathrm{m}_{A} = 0.80\pm0.05 \msun$. 

We used the parallel-tempering MCMC sampler from \emcee \citep{Foreman-Mackey2013}. We used 10 temperatures with the default temperature scale, 300 walkers, and iterated for $10^5$ steps. We examined the walker position plots, autocorrelation functions, and the evolution of the median and 1$\sigma$ values of each parameter to assess when the burn-in phase was complete. We present the constraints on the mass, eccentricity, and other orbital parameters in column (2) of Table \ref{tab:orb_results}. 

%%%%%%%%%%%%%%%%%%%%%%%%%%%%%%%%%
\subsection{White Dwarf Cooling Models}
\label{subsection:wd_models}

Our initial MCMC analysis leaves out one additional source of information on the mass and orbit of HD 159062 B: the photometric constraints from the ShaneAO, PHARO, and NIRC2 images. In combination with white dwarf cooling models, these data can provide new information about the mass and age of the white dwarf, in turn informing the orbital parameters.

We therefore implemented a second MCMC analysis with an additional prior that made use of white dwarf cooling models to constrain the mass and cooling age of the companion using the photometry in $J$, \ks\ and \lp. For this prior, we included the cooling age of the white dwarf as a free parameter in our model. 

We used the Montreal white dwarf cooling models, presented as a function of the white dwarf mass, from \citet{Holberg2006,Kowalski2006,Tremblay2011,Bergeron2011}\footnote{Available online at  \url{http://www.astro.umontreal.ca/~bergeron/CoolingModels}}. These models provide synthetic absolute magnitudes in a variety of standard filters, including $J$ and \ks. \lp\ synthetic absolute magnitudes were provided by P. Bergeron in private communication, since they were not available in the default model grids. For this analysis, we used the pure hydrogen atmosphere models.

We note that our choice of the pure hydrogen atmosphere models was motivated primarily by the commonality of the DA spectral type. However, \citet{Kowalski2006} have determined via opacity modeling that the majority of cool white dwarfs probably have hydrogen-dominated atmospheres, so this choice is well-justified. Additionally, \citet{Bergeron2019} analyzed the masses of new white dwarf candidates from Gaia, and demonstrated that the pure-hydrogen assumption yields more realistic mass determinations for the coolest white dwarfs with $\teff<5000$K. They predict that cool white dwarfs are most likely to be of the hydrogen-dominated DC type. Finally, the location of the companion in the  $J-\ks$ vs. $\ks-\lp$ color-color space is more consistent with the pure hydrogen models than the pure helium models. 

For each step in the MCMC chain, and for each of the three filters, we used the model mass and cooling age to interpolate model absolute magnitudes from the cooling curves. We used the parallax and known apparent magnitudes of the primary star to calculate the model \dm\ values in each filter. For this step, we again used WISE $W1$ as a proxy for \lp\ for HD 159062 A, since no precise \lp\ photometry was readily available. We then applied a Gaussian prior on the derived model \dm, with median $\mu$ and standard deviation $\sigma$ taken from the observed contrast and uncertainty from the ShaneAO, PHARO, and NIRC2 images, $\Delta J = 10.09\pm0.38$ mag., $\Delta \ks = 10.06\pm0.22$ mag., and $\Delta \lp = 9.67\pm0.08$ mag. We additionally constrained the cooling age to be $\tau <13.8$ Gyr, the age of the universe.

The constraints from the combined dynamical and photometric analysis are reported in column (3) of Table \ref{tab:orb_results}. 

\begin{table}[t]
    \centering
    \caption{MCMC Results}
    \begin{tabular}{lcc}
        \hline
        {\bf Parameter} & \multicolumn{2}{c}{\bf Median \& 68\% CI} \\
        \hline
        Model & RV/Ast & RV/Ast/Phot \\
        \hline
        %\multicolumn{3}{l}{Sample Parameters} \\
        %\hline
        m$_{A}$ (\msun) & $0.80\pm0.05$ & $0.80\pm0.05$ \\
        m$_{B}$ (\msun) & $0.65^{+0.14}_{-0.04}$ & $0.65^{+0.12}_{-0.04}$ \\
        $\log{\mathrm{P}}$ (yr) & $2.38^{+0.19}_{-0.15}$ & $2.40^{+0.18}_{-0.16}$ \\
        $\sqrt{e}\cos{\omega}$ & $0.61^{+0.16}_{-0.42}$ & $0.58^{+0.18}_{-0.46}$ \\
        $\sqrt{e}\sin{\omega}$ & $-0.27^{+0.07}_{-0.10}$ & $-0.27^{+0.07}_{-0.07}$ \\
        $\Omega$ ($^{\circ}$) & $138^{+11}_{-5}$ & $137^{+9}_{-4}$ \\
        $M_{0}(t_{i=0})$ ($^{\circ}$) & $144^{+59}_{-28}$ & $147^{+65}_{-27}$ \\
        $\cos{i}$ & $0.60^{+0.23}_{-0.13}$ & $0.58^{+0.22}_{-0.12}$ \\
        $\pi$ (mas) & $46.12\pm0.02$ & $46.12\pm0.02$ \\
        $\tau$ (Gyr) & \nodata & $8.2^{+0.3}_{-0.5}$ \\
        \hline
        \multicolumn{3}{l}{Derived Parameters} \\
        \hline
        P (yr) & $238^{+128}_{-68}$ & $250^{+130}_{-76}$ \\
        $e$ & $0.44^{+0.30}_{-0.31}$ & $0.40^{+0.31}_{-0.28}$ \\
        $\omega$ ($^{\circ}$) & $-26^{+7}_{-29}$ & $-26^{+7}_{-38}$ \\
        $i$ ($^{\circ}$) & $53^{+9}_{-19}$ & $54^{+8}_{-18}$ \\
        $T_{eff\mathrm{,B}}$ (K) & \nodata & $4580^{+440}_{-160}$ \\
        \hline
        \multicolumn{3}{l}{Instrumental Parameters} \\
        \hline        
        $\gamma_k$ (\ms) & $934^{+492}_{-605}$ & $1018^{+431}_{-572}$ \\
        $\gamma_j$ (\ms) & $933^{+492}_{-605}$ & $1018^{+431}_{-573}$ \\
        $\sigma_{jit,k}$ (\ms) & $2.3^{+2.9}_{-1.5}$ & $2.3^{+2.9}_{-1.5}$ \\
        $\sigma_{jit,j}$ (\ms) & $1.35^{+0.33}_{-0.30}$ & $1.36^{+0.32}_{-0.30}$ \\
    \end{tabular}
    \label{tab:orb_results}
\end{table}

\subsection{Results}

We find that all of the constraints from the combined analysis are consistent to within 1$\sigma$ with those from the RV and astrometric analysis. The only additional constraints resulting from inclusion of the photometric prior are on the cooling age and temperature of the white dwarf. We find that the photometry is consistent with an old white dwarf ($\tau = 8.2^{+0.3}_{-0.5}$ Gyr), which has cooled significantly.

\begin{figure*}
    \centering
    \includegraphics[width=\textwidth]{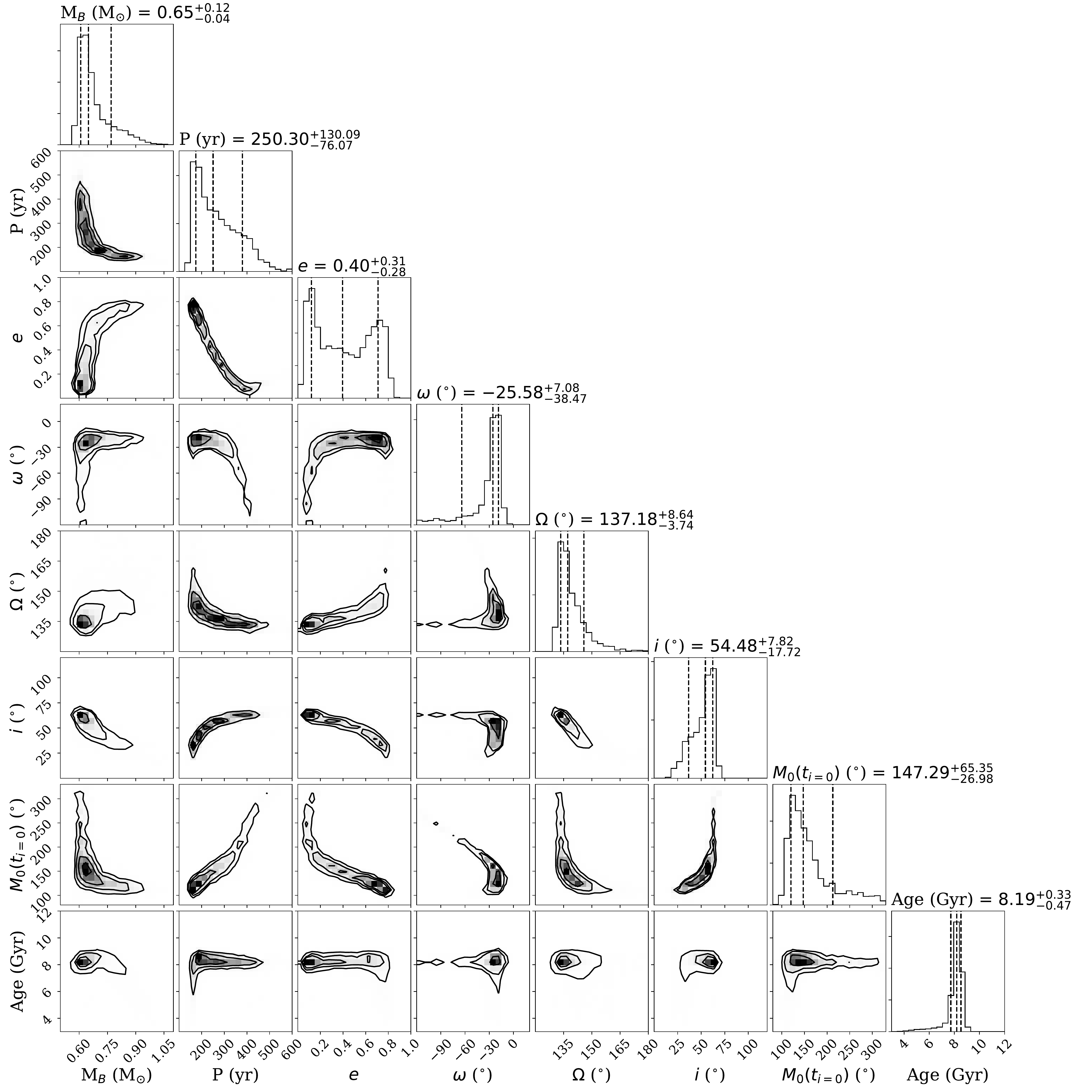}
    \caption{Corner plot of the derived Keplerian orbital elements, mass, and age of HD 159062 B from the joint RV, astrometric, and photometric MCMC analysis. The posteriors are evidently highly covariant. Correlations between the period, mass, inclination, and eccentricity are most likely due to the short time coverage of the RV and astrometric data used in the MCMC analysis, relative to the long orbital period of the binary.}
    \label{fig:corner}
\end{figure*}

To demonstrate the full behavior of the posteriors, including correlations between the various orbital elements, mass, and age, a corner plot of the derived white dwarf parameters is displayed in Figure \ref{fig:corner}. It is clear from the covariance plots that many of the posteriors on orbital parameters are highly correlated, not surprising given the limited coverage of the full orbit available in the RV and imaging data.

The white dwarf companion HD 159062 B has a mass constraint of $m_{B} = 0.65^{+0.12}_{-0.04}$ \msun\ and a long orbital period of $P = 250^{+130}_{-76}$ years, much longer than the 14 year baseline of our RV observations. The period remains poorly constrained, with a $1\sigma$ credible interval ranging from 174 -- 380 years. The eccentricity posterior appears to be multi-modal, with a peak at low eccentricity near $e=0.1$ and another at high eccentricity near $e=0.7$.

\begin{figure*}
    \centering
    \includegraphics[width=\textwidth]{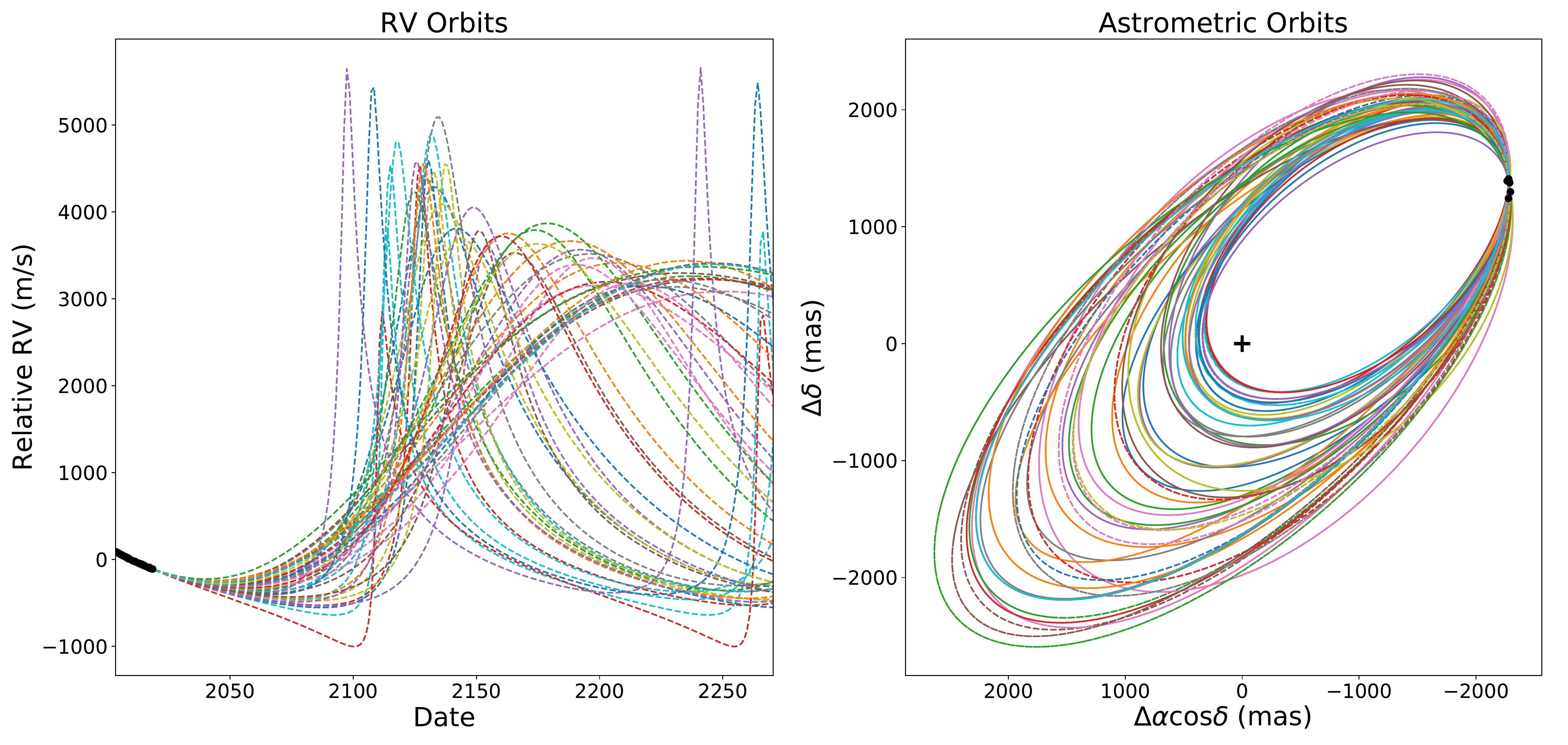}
    \caption{A random sample of 50  orbits drawn from the posterior probability distributions for HD 159062 B. This plot demonstrates that the orbital parameters are not fully constrained by the limited number of both RV and especially astrometric data points available. Due to the very long predicted orbital periods (typically hundreds of years), it will take the detection of astrometric orbital motion before the orbital parameters can be more precisely pinned down.}
    \label{fig:orbits}
\end{figure*}

A plot showing a random sampling of 50 orbits drawn from the posteriors of the joint MCMC run is provided in Figure \ref{fig:orbits}. From this plot, it is clear that the orbital period is not fully constrained with the data available. 

The two eccentricity modes can be seen in the possible radial velocity curves, with some solutions showing sharp peaks indicative of a high eccentricity, while others are more sinusoidal in shape. However, both classes of orbit have the same astrometric signature, due to an anti-correlation between eccentricity and inclination. As might be expected, the low eccentricity solutions have more edge-on orientations while the higher eccentricity solutions are closer to face on, such that both types of orbit still fit the astrometric motion well. Additionally, eccentricity is highly anti-correlated with orbital period, with shorter orbital periods corresponding to higher eccentricities. A correlation between mass and period is also observed, so that the range of possible orbits follows a continuum from lower-mass, shorter-period, circular, edge-on cases to higher-mass, longer-period, eccentric, face-on solutions.

\begin{figure}
    \centering
    \includegraphics[width=0.5\textwidth]{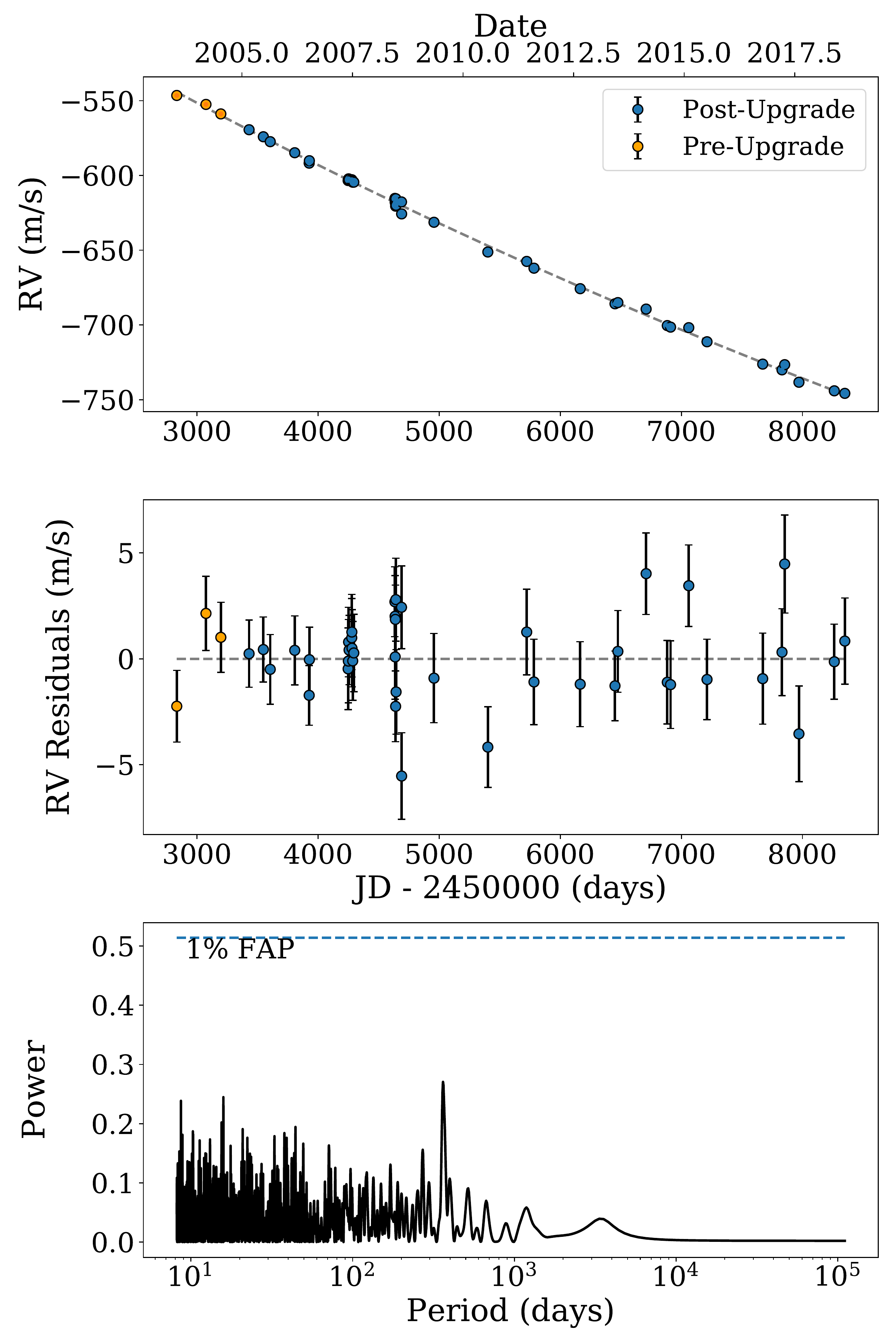}
    \caption{{\it Top}: Maximum probability orbit plotted against the radial velocity data. {\it Middle}: Residuals of radial velocities after maximum probability orbit was subtracted. We test for additional orbit signatures in these residuals but do not find any significant peaks. {\it Bottom}: Lomb Scargle periodogram of RV residuals. No significant peaks are evident in the periodogram, implying that the data do not detect any inner planetary or stellar companions to HD 159062\ A. The 1\% empirical false alarm probability, calculated as in \citet{Howard2016}, is plotted for reference.}
    \label{fig:rv_resids}
\end{figure}

The maximum probability orbit model is plotted against the RV data in Figure \ref{fig:rv_resids}, with residuals shown. A periodogram of these residuals is also displayed. Together, these plots demonstrate that the RV trend and slight curvature are fully accounted for in the single-companion Keplerian orbital fit. The residuals do not show correlated structure, and no significant shorter-period peaks are seen in the periodogram.

Using the posteriors on mass and cooling age and the Montreal white dwarf cooling models, we calculate the posterior on the temperature of the white dwarf as well. This can be compared to the \teff\ implied by the white dwarf spectrum if spectroscopy is obtained for this target. We obtain a derived constraint on HD 159062 B's temperature of $\teff = 4580^{+440}_{-160}$ K, slightly cooler than the main sequence primary star, HD 159062 A. 

\begin{figure}
    \centering
    \includegraphics[width=0.5\textwidth]{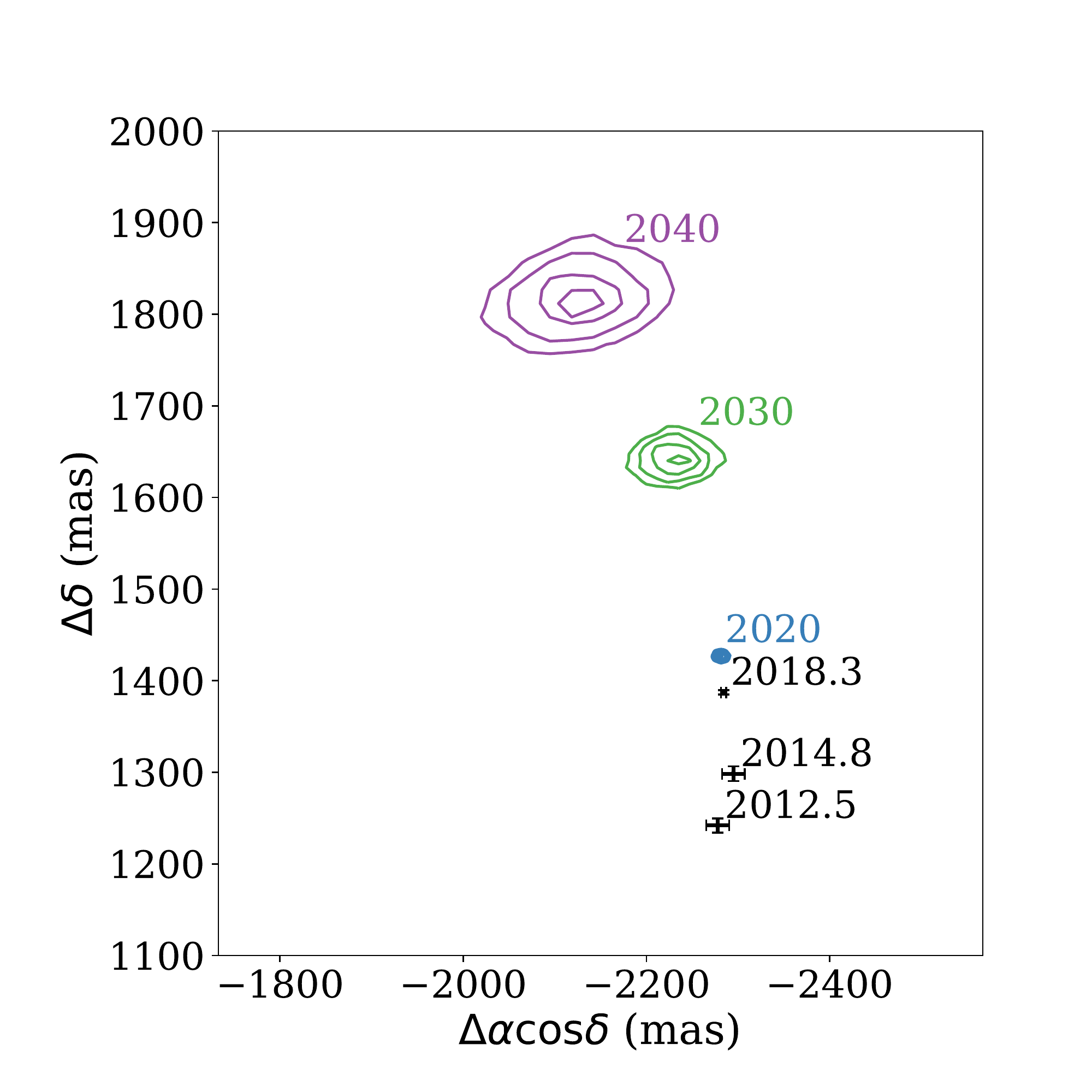}
    \caption{Prediction of the future location of the white dwarf. The posterior probability distributions on the orbital parameters were used to produce posteriors on the future position of the companion for dates in the next two decades, as denoted in the plot. As the dates progress further into the future, the white dwarf's position becomes less certain.}
    \label{fig:future_astrom}
\end{figure}

Finally, for future observations, we predict the expected location of the white dwarf over the next two decades. We convert the posteriors on orbital parameters into posteriors on the angular offsets between HD 159062 A and B at several epochs spanning 2020 through 2040, and plot these 2D posterior distributions along with the astrometric data in Figure \ref{fig:future_astrom}. The plot demonstrates that orbital motion on the scale of a few hundred mas will be detectable in the next decade. As expected, the uncertainty on the position of the white dwarf increases over time due to uncertainties in the orbital parameters. More astrometric data from Keck/NIRC2 will help to place new constraints on the orbital parameters after several years.

\subsection{Total system age}
The combined mass and age constraints from the MCMC analysis can be used to assess the full age of the system.

We first use several published empirical initial-to-final mass relations for white dwarfs to determine the mass of the white dwarf progenitor. 
Two empirical linear relations between progenitor and final white dwarf mass are presented by \citet{Catalan2008}: The first uses all white dwarfs in their sample (their Equation 1), and yields an initial mass estimate for HD 159062 B of $m_{B,i} = 2.27^{+1.03}_{-0.36}$ \msun. The second fits low-mass and high-mass progenitors separately, with a division at an initial mass of 2.7\msun. Using the low-mass relation (their Equation 3), we derive a consistent initial mass of $m_{B,i} = 2.30^{+1.27}_{-0.46} \msun$. Others use a slightly different method to estimate the IFMR for white dwarfs in clusters of known age, by binning all stars in each cluster to obtain higher signal-to-noise. The relations provided in some of these works yield initial mass estimates of $2.35^{+1.13}_{-0.46}$ \msun\ \citep{Kalirai2008}  and $2.41^{+0.94}_{-0.34}$ \msun\ \citep{Williams2009} respectively, all in good agreement.

For progenitor masses of $\sim 2.4\msun$, the expected main sequence lifetime is approximately 7.6 Gyr based on the MIST evolutionary sequences \citep{Choi2016}, and this would make the full lifetime of the system ($t_{\rm MS}+t_{\rm cool}$) longer than the age of the universe. However, due to the significant  scatter in the initial-final mass relations for white dwarfs used here, this is not a cause for great concern. Reversing the question, we can impose an upper limit on the main sequence lifetime of the white dwarf progenitor of approximately 5.5 Gyr. This implies a lower limit on the progenitor's initial mass of $\gtrapprox 2.7$\msun, within the upper $1\sigma$ uncertainty range on all initial mass estimates reported here.

%%%%%%%%%%%%%%%%%%%%%%%%%%%%%%%%%%%%%%%%%
\section{Discussion}
\label{section:discussion}

The indirect suggestion of a white dwarf companion to HD 159062 A has been detailed in \citet{Fuhrmann2017,Fuhrmann2017b}. Specifically, the overabundance of Barium measured by these studies indicates contamination from an AGB companion sometime during the evolution of HD 159062 A. We do note, however, that other abundance surveys of the local neighborhood by \citet{Reddy2006,Mishenina2013} find lower and more consistent [Ba/Fe] abundance values of +0.17 and +0.15 dex respectively.

In comparison with typical Ba or CH stars, HD 159062 A differs in several ways. First, its \logg\ measured from high resolution spectroscopy indicates that HD 159062 A is a main sequence star. Typical Ba stars are observed to be G and K giants \citep{Pols2003,Izzard2010,Mahanta2016} although dwarf Ba systems do exist. CH stars are more often main sequence companions, but the metallicity of HD 159062 is too high to qualify as a typical CH star, which have $\feh <-0.5$ \citep{Escorza2019}. Ba binaries typically have orbital periods ranging from a few to a few tens of years. The observed separation of HD 159062 B makes it significantly wider than known Ba or CH star binaries, and its orbital period is $P = 250^{+130}_{-76}$ years, several times longer than typical Ba or CH binary orbital periods \citep{Jorissen1998,Jorissen2016,VanderSwaelmen2017,Escorza2019}.

The wide separation of HD 159062 B raises some doubts about its capability of contaminating HD 159062 A sufficiently via a stellar wind to explain the observed Ba abundance. \citet{Izzard2010} propose that the eccentricity-orbital period relation for Ba star binaries could be better explained by a process which ``kicks'' the white dwarf as it forms, possibly due to asymmetric mass loss or magnetic fields. In their model, a tail of high eccentricity, very long period Ba binaries is formed, possibly consistent with the orbit of HD 159062 B.

\section{Conclusions}
HD 159062 is a newly detected binary system with one old main sequence component, and one initially more massive component that has since evolved into a white dwarf. We have shown that HD 159062 B cannot be a main sequence or brown dwarf companion, and is also inconsistent with a background source. Based on its mass and photometry, it is consistent with a white dwarf companion. The system's old age means that the white dwarf has cooled significantly, with a temperature of $\teff{_{B}} = 4580{^{+440}}{_{-160}}$ K and cooling age of $t{_{\rm cool}} = 8.2{^{+0.3}}{_{-0.5}}$ Gyr. Continued monitoring of the radial velocities of HD 159062 A via high-resolution spectroscopy, and especially additional measurements of the astrometric position of HD 159062 B through high-contrast imaging will help to further constrain the long-period binary orbit and improve our understanding of the evolutionary history of this system.

\acknowledgements
We would like to thank the anonymous referee for many helpful and detailed suggestions that have helped to improve this manuscript. We thank Pierre Bergeron for providing updated white dwarf cooling models including \lp-band synthetic photometry. L.A.H. would like to thank Rob De Rosa, Eric Nielsen, James Graham, and Bruce Macintosh for helpful discussion on this project, and Erik Petigura for carrying out HIRES observations and performing a spectroscopic analysis of the properties of HD 159062 A. M.R.K acknowledges support from the NSF Graduate Research Fellowship, grant No. DGE 1339067. Some of the data presented herein were obtained at the W. M. Keck Observatory, which is operated as a scientific partnership among the California Institute of Technology, the University of California and the National Aeronautics and Space Administration. The Observatory was made possible by the generous financial support of the W. M. Keck Foundation. The authors wish to recognize and acknowledge the very significant cultural role and reverence that the summit of Maunakea has always had within the indigenous Hawaiian community.  We are most fortunate to have the opportunity to conduct observations from this mountain.

\facility{Keck:I (HIRES), Keck:II (NIRC2), Lick:Shane (ShARCS), Palomar:Hale (PHARO)}
\software{\emcee\citep{Foreman-Mackey2013}, \texttt{VIP} \citep{Gomez2017}, \texttt{isoclassify} \citep{Huber2017}, \texttt{pyKLIP} \citep{jasonwang2015}, \texttt{photutils} \citep{bradley2017}}

\bibliography{allbib}
\end{document}

%% file: rvs.tex
2452832.91668 & 89.22 & 1.62 & \nodata \\
2453074.09155 & 83.23 & 1.68 & \nodata \\
2453196.81631 & 76.88 & 1.58 & \nodata \\
2453430.15779 & 65.68 & 0.93 & 0.166 \\
2453547.94513 & 60.98 & 0.84 & 0.167 \\
2453604.84622 & 57.70 & 1.03 & 0.168 \\
2453807.13748 & 50.30 & 1.00 & 0.168 \\
2453926.45063 & 43.33 & 0.57 & 0.168 \\
2453927.88530 & 44.96 & 0.83 & 0.167 \\
2454248.03230 & 31.71 & 1.44 & 0.170 \\
2454249.95458 & 32.00 & 1.49 & 0.170 \\
2454252.04380 & 32.81 & 1.02 & 0.172 \\
2454255.93510 & 32.28 & 1.01 & 0.172 \\
2454277.86337 & 31.99 & 1.33 & 0.171 \\
2454278.90748 & 32.22 & 1.22 & 0.171 \\
2454279.94547 & 31.42 & 1.29 & 0.164 \\
2454285.90800 & 30.58 & 1.34 & 0.170 \\
2454294.90713 & 30.60 & 1.30 & 0.173 \\
2454634.43562 & 19.74 & 1.03 & 0.170 \\
2454635.94255 & 19.00 & 1.43 & 0.170 \\
2454636.93316 & 17.04 & 1.53 & 0.171 \\
2454638.33247 & 18.76 & 0.99 & 0.171 \\
2454640.38633 & 14.57 & 1.06 & 0.171 \\
2454641.94950 & 19.55 & 1.47 & 0.172 \\
2454644.08325 & 15.11 & 1.52 & 0.170 \\
2454688.88580 & 17.38 & 1.47 & 0.175 \\
2454689.89920 & 9.37 & 1.58 & 0.174 \\
2454957.13845 & 3.77 & 1.67 & 0.178 \\
2455401.78282 & -16.12 & 1.40 & 0.172 \\
2455722.86450 & -22.42 & 1.56 & 0.176 \\
2455782.89029 & -26.95 & 1.55 & 0.170 \\
2456164.72278 & -40.66 & 1.54 & 0.172 \\
2456451.97582 & -50.77 & 1.03 & 0.169 \\
2456475.78721 & -49.97 & 1.43 & 0.173 \\
2456709.08945 & -54.30 & 1.43 & 0.172 \\
2456883.74892 & -65.35 & 1.49 & 0.171 \\
2456910.83550 & -66.37 & 1.62 & 0.171 \\
2457061.17682 & -66.74 & 1.43 & 0.169 \\
2457211.95513 & -76.17 & 1.40 & 0.170 \\
2457671.70374 & -91.12 & 1.72 & 0.167 \\
2457831.16397 & -94.97 & 1.59 & 0.167 \\
2457853.14530 & -91.51 & 1.92 & 0.170 \\
2457970.94501 & -103.25 & 1.86 & 0.169 \\
2458263.03029 & -108.99 & 1.22 & 0.168 \\
2458349.72648 & -110.69 & 1.58 & 0.167 \\

%% file: main.bbl
\begin{thebibliography}{}
\expandafter\ifx\csname natexlab\endcsname\relax\def\natexlab#1{#1}\fi

\bibitem[{{Bergeron} {et~al.}(2019){Bergeron}, {Dufour}, {Fontaine}, {Coutu},
  {Blouin}, {Genest-Beaulieu}, {B{\'e}dard}, \& {Rolland}}]{Bergeron2019}
{Bergeron}, P., {Dufour}, P., {Fontaine}, G., {et~al.} 2019, arXiv e-prints,
  arXiv:1904.02022

\bibitem[{{Bergeron} {et~al.}(2011){Bergeron}, {Wesemael}, {Dufour},
  {Beauchamp}, {Hunter}, {Saffer}, {Gianninas}, {Ruiz}, {Limoges}, {Dufour},
  {Fontaine}, \& {Liebert}}]{Bergeron2011}
{Bergeron}, P., {Wesemael}, F., {Dufour}, P., {et~al.} 2011, \apj, 737, 28

\bibitem[{{Boffin} \& {Jorissen}(1988)}]{Boffin1988}
{Boffin}, H.~M.~J., \& {Jorissen}, A. 1988, \aap, 205, 155

\bibitem[{Bradley {et~al.}(2017)Bradley, Sipocz, Robitaille, Vinícius,
  Tollerud, Deil, Barbary, Günther, Cara, Busko, Droettboom, Bostroem, Bray,
  Bratholm, Pickering, Craig, Barentsen, Pascual, Conseil, adonath, Greco,
  Kerzendorf, de~Val-Borro, StuartLittlefair, Ogaz, Lim, Ferreira, D'Eugenio,
  \& Weaver}]{bradley2017}
Bradley, L., Sipocz, B., Robitaille, T., {et~al.} 2017, astropy/photutils:
  v0.4, , , doi:10.5281/zenodo.1039309

\bibitem[{{Brewer} {et~al.}(2016){Brewer}, {Fischer}, {Valenti}, \&
  {Piskunov}}]{Brewer2016}
{Brewer}, J.~M., {Fischer}, D.~A., {Valenti}, J.~A., \& {Piskunov}, N. 2016,
  The Astrophysical Journal Supplement Series, 225, 32

\bibitem[{{Casagrande} {et~al.}(2011){Casagrande}, {Sch{\"o}nrich}, {Asplund},
  {Cassisi}, {Ram{\'\i}rez}, {Mel{\'e}ndez}, {Bensby}, \&
  {Feltzing}}]{Casagrande2011}
{Casagrande}, L., {Sch{\"o}nrich}, R., {Asplund}, M., {et~al.} 2011, \aap, 530,
  A138

\bibitem[{{Catal{\'a}n} {et~al.}(2008){Catal{\'a}n}, {Isern},
  {Garc{\'{\i}}a-Berro}, \& {Ribas}}]{Catalan2008}
{Catal{\'a}n}, S., {Isern}, J., {Garc{\'{\i}}a-Berro}, E., \& {Ribas}, I. 2008,
  \mnras, 387, 1693

\bibitem[{{Choi} {et~al.}(2016){Choi}, {Dotter}, {Conroy}, {Cantiello},
  {Paxton}, \& {Johnson}}]{Choi2016}
{Choi}, J., {Dotter}, A., {Conroy}, C., {et~al.} 2016, \apj, 823, 102

\bibitem[{Crepp {et~al.}(2013)Crepp, Johnson, Howard, Marcy, Gianninas, Kilic,
  \& Wright}]{Crepp2013a}
Crepp, J.~R., Johnson, J.~A., Howard, A.~W., {et~al.} 2013, The Astrophysical
  Journal, 774, 1

\bibitem[{Crepp {et~al.}(2012)Crepp, Johnson, Howard, Marcy, Fischer,
  Hillenbrand, Yantek, Delaney, Wright, Isaacson, \& Montet}]{Crepp2012}
---. 2012, The Astrophysical Journal, 761, 39

\bibitem[{{Crepp} {et~al.}(2018){Crepp}, {Gonzales}, {Bowler}, {Morales},
  {Stone}, {Spalding}, {Vaz}, {Hinz}, {Ertel}, {Howard}, \&
  {Isaacson}}]{Crepp2018}
{Crepp}, J.~R., {Gonzales}, E.~J., {Bowler}, B.~P., {et~al.} 2018, \apj, 864,
  42

\bibitem[{{Cutri} {et~al.}(2003){Cutri}, {Skrutskie}, {van Dyk}, {Beichman},
  {Carpenter}, {Chester}, {Cambresy}, {Evans}, {Fowler}, {Gizis}, {Howard},
  {Huchra}, {Jarrett}, {Kopan}, {Kirkpatrick}, {Light}, {Marsh}, {McCallon},
  {Schneider}, {Stiening}, {Sykes}, {Weinberg}, {Wheaton}, {Wheelock}, \&
  {Zacarias}}]{Cutri2003}
{Cutri}, R.~M., {Skrutskie}, M.~F., {van Dyk}, S., {et~al.} 2003, {2MASS All
  Sky Catalog of point sources.}

\bibitem[{{De Rosa} {et~al.}(2016){De Rosa}, {Rameau}, {Patience}, {Graham},
  {Doyon}, {Lafreni{\`e}re}, {Macintosh}, {Pueyo}, {Rajan}, {Wang},
  {Ward-Duong}, {Hung}, {Maire}, {Nielsen}, {Ammons}, {Bulger}, {Cardwell},
  {Chilcote}, {Galvez}, {Gerard}, {Goodsell}, {Hartung}, {Hibon}, {Ingraham},
  {Johnson-Groh}, {Kalas}, {Konopacky}, {Marchis}, {Marois}, {Metchev},
  {Morzinski}, {Oppenheimer}, {Perrin}, {Rantakyr{\"o}}, {Savransky}, \&
  {Thomas}}]{DeRosa2016}
{De Rosa}, R.~J., {Rameau}, J., {Patience}, J., {et~al.} 2016, \apj, 824, 121

\bibitem[{{Escorza} {et~al.}(2019){Escorza}, {Karinkuzhi}, {Jorissen}, {Siess},
  {Van Winckel}, {Pourbaix}, {Johnston}, {Miszalski}, {Oomen}, {Abdul-Masih},
  {Boffin}, {North}, {Manick}, {Shetye}, \& {Miko{\l}ajewska}}]{Escorza2019}
{Escorza}, A., {Karinkuzhi}, D., {Jorissen}, A., {et~al.} 2019, arXiv e-prints,
  arXiv:1904.04095

\bibitem[{{Foreman-Mackey} {et~al.}(2013){Foreman-Mackey}, {Hogg}, {Lang}, \&
  {Goodman}}]{Foreman-Mackey2013}
{Foreman-Mackey}, D., {Hogg}, D.~W., {Lang}, D., \& {Goodman}, J. 2013, \pasp,
  125, 306

\bibitem[{{Fuhrmann} {et~al.}(2017{\natexlab{a}}){Fuhrmann}, {Chini},
  {Kaderhandt}, \& {Chen}}]{Fuhrmann2017}
{Fuhrmann}, K., {Chini}, R., {Kaderhandt}, L., \& {Chen}, Z.
  2017{\natexlab{a}}, \mnras, 464, 2610

\bibitem[{{Fuhrmann} {et~al.}(2017{\natexlab{b}}){Fuhrmann}, {Chini},
  {Kaderhandt}, {Chen}, \& {Lachaume}}]{Fuhrmann2017b}
{Fuhrmann}, K., {Chini}, R., {Kaderhandt}, L., {Chen}, Z., \& {Lachaume}, R.
  2017{\natexlab{b}}, \mnras, 471, 3768

\bibitem[{{Gagn{\'e}} {et~al.}(2018){Gagn{\'e}}, {Mamajek}, {Malo}, {Riedel},
  {Rodriguez}, {Lafreni{\`e}re}, {Faherty}, {Roy-Loubier}, {Pueyo}, {Robin}, \&
  {Doyon}}]{Gagne2018}
{Gagn{\'e}}, J., {Mamajek}, E.~E., {Malo}, L., {et~al.} 2018, \apj, 856, 23

\bibitem[{{Gaia Collaboration} {et~al.}(2018){Gaia Collaboration}, {Brown},
  {Vallenari}, {Prusti}, {de Bruijne}, {Babusiaux}, {Bailer-Jones}, {Biermann},
  {Evans}, {Eyer}, \& et~al.}]{Gaia2018}
{Gaia Collaboration}, {Brown}, A.~G.~A., {Vallenari}, A., {et~al.} 2018, \aap,
  616, A1

\bibitem[{{Gavel} {et~al.}(2014){Gavel}, {Kupke}, {Dillon}, {Norton},
  {Ratliff}, {Cabak}, {Phillips}, {Rockosi}, {McGurk}, {Srinath}, {Peck},
  {Deich}, {Lanclos}, {Gates}, {Saylor}, {Ward}, \& {Pfister}}]{Gavel2014}
{Gavel}, D., {Kupke}, R., {Dillon}, D., {et~al.} 2014, in \procspie, Vol. 9148,
  Adaptive Optics Systems IV, 914805

\bibitem[{{Giammichele} {et~al.}(2012){Giammichele}, {Bergeron}, \&
  {Dufour}}]{Giammichele2012}
{Giammichele}, N., {Bergeron}, P., \& {Dufour}, P. 2012, \apjs, 199, 29

\bibitem[{{Gomez Gonzalez} {et~al.}(2017){Gomez Gonzalez}, {Wertz}, {Absil},
  {Christiaens}, {Defr{\`e}re}, {Mawet}, {Milli}, {Absil}, {Van Droogenbroeck},
  {Cantalloube}, {Hinz}, {Skemer}, {Karlsson}, \& {Surdej}}]{Gomez2017}
{Gomez Gonzalez}, C.~A., {Wertz}, O., {Absil}, O., {et~al.} 2017, \aj, 154, 7

\bibitem[{{Hayward} {et~al.}(2001){Hayward}, {Brandl}, {Pirger}, {Blacken},
  {Gull}, {Schoenwald}, \& {Houck}}]{Hayward2001}
{Hayward}, T.~L., {Brandl}, B., {Pirger}, B., {et~al.} 2001, \pasp, 113, 105

\bibitem[{{Holberg} \& {Bergeron}(2006)}]{Holberg2006}
{Holberg}, J.~B., \& {Bergeron}, P. 2006, \aj, 132, 1221

\bibitem[{{Holberg} {et~al.}(2016){Holberg}, {Oswalt}, {Sion}, \&
  {McCook}}]{Holberg2016}
{Holberg}, J.~B., {Oswalt}, T.~D., {Sion}, E.~M., \& {McCook}, G.~P. 2016,
  \mnras, 462, 2295

\bibitem[{{Hollands} {et~al.}(2018){Hollands}, {Tremblay}, {G{\"a}nsicke},
  {Gentile-Fusillo}, \& {Toonen}}]{Hollands2018}
{Hollands}, M.~A., {Tremblay}, P.-E., {G{\"a}nsicke}, B.~T., {Gentile-Fusillo},
  N.~P., \& {Toonen}, S. 2018, \mnras, 480, 3942

\bibitem[{{Howard} \& {Fulton}(2016)}]{Howard2016}
{Howard}, A.~W., \& {Fulton}, B.~J. 2016, \pasp, 128, 114401

\bibitem[{Howard {et~al.}(2010)Howard, Marcy, Johnson, Fischer, Wright,
  Isaacson, Valenti, Anderson, Lin, \& Ida}]{Howard2010}
Howard, A.~W., Marcy, G.~W., Johnson, J.~A., {et~al.} 2010, Science, 330, 653

\bibitem[{{Huber} {et~al.}(2017){Huber}, {Zinn}, {Bojsen-Hansen},
  {Pinsonneault}, {Sahlholdt}, {Serenelli}, {Silva Aguirre}, {Stassun},
  {Stello}, {Tayar}, {Bastien}, {Bedding}, {Buchhave}, {Chaplin}, {Davies},
  {Garc{\'{\i}}a}, {Latham}, {Mathur}, {Mosser}, \& {Sharma}}]{Huber2017}
{Huber}, D., {Zinn}, J., {Bojsen-Hansen}, M., {et~al.} 2017, \apj, 844, 102

\bibitem[{{Huby} {et~al.}(2017){Huby}, {Bottom}, {Femenia}, {Ngo}, {Mawet},
  {Serabyn}, \& {Absil}}]{Huby2017}
{Huby}, E., {Bottom}, M., {Femenia}, B., {et~al.} 2017, \aap, 600, A46

\bibitem[{{Izzard} {et~al.}(2010){Izzard}, {Dermine}, \& {Church}}]{Izzard2010}
{Izzard}, R.~G., {Dermine}, T., \& {Church}, R.~P. 2010, \aap, 523, A10

\bibitem[{{Jim{\'e}nez-Esteban} {et~al.}(2018){Jim{\'e}nez-Esteban}, {Torres},
  {Rebassa-Mansergas}, {Skorobogatov}, {Solano}, {Cantero}, \&
  {Rodrigo}}]{Jimenez-Esteban2018}
{Jim{\'e}nez-Esteban}, F.~M., {Torres}, S., {Rebassa-Mansergas}, A., {et~al.}
  2018, \mnras, 480, 4505

\bibitem[{{Jorissen} {et~al.}(1998){Jorissen}, {Van Eck}, {Mayor}, \&
  {Udry}}]{Jorissen1998}
{Jorissen}, A., {Van Eck}, S., {Mayor}, M., \& {Udry}, S. 1998, \aap, 332, 877

\bibitem[{{Jorissen} {et~al.}(2016){Jorissen}, {Van Eck}, {Van Winckel},
  {Merle}, {Boffin}, {Andersen}, {Nordstr{\"o}m}, {Udry}, {Masseron},
  {Lenaerts}, \& {Waelkens}}]{Jorissen2016}
{Jorissen}, A., {Van Eck}, S., {Van Winckel}, H., {et~al.} 2016, \aap, 586,
  A158

\bibitem[{{Kalirai} {et~al.}(2008){Kalirai}, {Hansen}, {Kelson}, {Reitzel},
  {Rich}, \& {Richer}}]{Kalirai2008}
{Kalirai}, J.~S., {Hansen}, B.~M.~S., {Kelson}, D.~D., {et~al.} 2008, \apj,
  676, 594

\bibitem[{{Kilic} {et~al.}(2018){Kilic}, {Hambly}, {Bergeron},
  {Genest-Beaulieu}, \& {Rowell}}]{Kilic2018}
{Kilic}, M., {Hambly}, N.~C., {Bergeron}, P., {Genest-Beaulieu}, C., \&
  {Rowell}, N. 2018, \mnras, 479, L113

\bibitem[{{Kowalski} \& {Saumon}(2006)}]{Kowalski2006}
{Kowalski}, P.~M., \& {Saumon}, D. 2006, \apjl, 651, L137

\bibitem[{{Leggett} {et~al.}(2001){Leggett}, {Allard}, {Geballe}, {Hauschildt},
  \& {Schweitzer}}]{Leggett2001}
{Leggett}, S.~K., {Allard}, F., {Geballe}, T.~R., {Hauschildt}, P.~H., \&
  {Schweitzer}, A. 2001, \apj, 548, 908

\bibitem[{{Mahanta} {et~al.}(2016){Mahanta}, {Karinkuzhi}, {Goswami}, \&
  {Duorah}}]{Mahanta2016}
{Mahanta}, U., {Karinkuzhi}, D., {Goswami}, A., \& {Duorah}, K. 2016, \mnras,
  463, 1213

\bibitem[{Mamajek \& Hillenbrand(2008)}]{Mamajek2008}
Mamajek, E.~E., \& Hillenbrand, L.~A. 2008, The Astrophysical Journal, 687,
  1264

\bibitem[{{Marois} {et~al.}(2006){Marois}, {Lafreni{\`e}re}, {Doyon},
  {Macintosh}, \& {Nadeau}}]{Marois2006}
{Marois}, C., {Lafreni{\`e}re}, D., {Doyon}, R., {Macintosh}, B., \& {Nadeau},
  D. 2006, \apj, 641, 556

\bibitem[{{Metchev}(2006)}]{Metchev2006}
{Metchev}, S.~A. 2006, PhD thesis, California Institute of Technology,
  California, USA

\bibitem[{{Mishenina} {et~al.}(2013){Mishenina}, {Pignatari}, {Korotin},
  {Soubiran}, {Charbonnel}, {Thielemann}, {Gorbaneva}, \&
  {Basak}}]{Mishenina2013}
{Mishenina}, T.~V., {Pignatari}, M., {Korotin}, S.~A., {et~al.} 2013, \aap,
  552, A128

\bibitem[{Mishenina {et~al.}(2008)Mishenina, Soubiran, Bienaym{\'{e}}, Korotin,
  Belik, Usenko, \& Kovtyukh}]{Mishenina2008}
Mishenina, T.~V., Soubiran, C., Bienaym{\'{e}}, O., {et~al.} 2008, Astronomy
  and Astrophysics, 489, 923

\bibitem[{{Morzinski} {et~al.}(2015){Morzinski}, {Males}, {Skemer}, {Close},
  {Hinz}, {Rodigas}, {Puglisi}, {Esposito}, {Riccardi}, {Pinna}, {Xompero},
  {Briguglio}, {Bailey}, {Follette}, {Kopon}, {Weinberger}, \&
  {Wu}}]{Morzinski2015}
{Morzinski}, K.~M., {Males}, J.~R., {Skemer}, A.~J., {et~al.} 2015, \apj, 815,
  108

\bibitem[{{Petigura} {et~al.}(2017){Petigura}, {Howard}, {Marcy}, {Johnson},
  {Isaacson}, {Cargile}, {Hebb}, {Fulton}, {Weiss}, {Morton}, {Winn}, {Rogers},
  {Sinukoff}, {Hirsch}, \& {Crossfield}}]{Petigura2017}
{Petigura}, E.~A., {Howard}, A.~W., {Marcy}, G.~W., {et~al.} 2017, \aj, 154,
  107

\bibitem[{{Piskunov} \& {Valenti}(2017)}]{Piskunov2017}
{Piskunov}, N., \& {Valenti}, J.~A. 2017, \aap, 597, A16

\bibitem[{{Pols} {et~al.}(2003){Pols}, {Karakas}, {Lattanzio}, \&
  {Tout}}]{Pols2003}
{Pols}, O.~R., {Karakas}, A.~I., {Lattanzio}, J.~C., \& {Tout}, C.~A. 2003, in
  Astronomical Society of the Pacific Conference Series, Vol. 303, Symbiotic
  Stars Probing Stellar Evolution, ed. R.~L.~M. {Corradi}, J.~{Mikolajewska},
  \& T.~J. {Mahoney}, 290

\bibitem[{{Ram{\'\i}rez} {et~al.}(2013){Ram{\'\i}rez}, {Allende Prieto}, \&
  {Lambert}}]{Ramirez2013}
{Ram{\'\i}rez}, I., {Allende Prieto}, C., \& {Lambert}, D.~L. 2013, \apj, 764,
  78

\bibitem[{{Ram{\'\i}rez} {et~al.}(2012){Ram{\'\i}rez}, {Fish}, {Lambert}, \&
  {Allende Prieto}}]{Ramirez2012}
{Ram{\'\i}rez}, I., {Fish}, J.~R., {Lambert}, D.~L., \& {Allende Prieto}, C.
  2012, \apj, 756, 46

\bibitem[{{Reddy} {et~al.}(2006){Reddy}, {Lambert}, \& {Allende
  Prieto}}]{Reddy2006}
{Reddy}, B.~E., {Lambert}, D.~L., \& {Allende Prieto}, C. 2006, \mnras, 367,
  1329

\bibitem[{{Serabyn} {et~al.}(2017){Serabyn}, {Huby}, {Matthews}, {Mawet},
  {Absil}, {Femenia}, {Wizinowich}, {Karlsson}, {Bottom}, {Campbell},
  {Carlomagno}, {Defr{\`e}re}, {Delacroix}, {Forsberg}, {Gomez Gonzalez},
  {Habraken}, {Jolivet}, {Liewer}, {Lilley}, {Piron}, {Reggiani}, {Surdej},
  {Tran}, {Vargas Catal{\'a}n}, \& {Wertz}}]{Serabyn2017}
{Serabyn}, E., {Huby}, E., {Matthews}, K., {et~al.} 2017, \aj, 153, 43

\bibitem[{{Service} {et~al.}(2016){Service}, {Lu}, {Campbell}, {Sitarski},
  {Ghez}, \& {Anderson}}]{Service2016}
{Service}, M., {Lu}, J.~R., {Campbell}, R., {et~al.} 2016, \pasp, 128, 095004

\bibitem[{{Sion} {et~al.}(2014){Sion}, {Holberg}, {Oswalt}, {McCook},
  {Wasatonic}, \& {Myszka}}]{Sion2014}
{Sion}, E.~M., {Holberg}, J.~B., {Oswalt}, T.~D., {et~al.} 2014, \aj, 147, 129

\bibitem[{{Soummer} {et~al.}(2012){Soummer}, {Pueyo}, \&
  {Larkin}}]{Soummer2012}
{Soummer}, R., {Pueyo}, L., \& {Larkin}, J. 2012, \apjl, 755, L28

\bibitem[{{Srinath} {et~al.}(2014){Srinath}, {McGurk}, {Rockosi}, {Kupke},
  {Gavel}, {Cabak}, {Cowley}, {Peck}, {Ratliff}, {Gates}, {Dillon}, {Norton},
  \& {Reining}}]{Srinath2014}
{Srinath}, S., {McGurk}, R., {Rockosi}, C., {et~al.} 2014, in \procspie, Vol.
  9148, Adaptive Optics Systems IV, 91482Z

\bibitem[{{Toonen} {et~al.}(2017){Toonen}, {Hollands}, {G{\"a}nsicke}, \&
  {Boekholt}}]{Toonen2017}
{Toonen}, S., {Hollands}, M., {G{\"a}nsicke}, B.~T., \& {Boekholt}, T. 2017,
  \aap, 602, A16

\bibitem[{{Tremblay} {et~al.}(2011){Tremblay}, {Bergeron}, \&
  {Gianninas}}]{Tremblay2011}
{Tremblay}, P.-E., {Bergeron}, P., \& {Gianninas}, A. 2011, \apj, 730, 128

\bibitem[{{van Dam} {et~al.}(2006){van Dam}, {Bouchez}, {Le Mignant},
  {Johansson}, {Wizinowich}, {Campbell}, {Chin}, {Hartman}, {Lafon}, {Stomski},
  \& {Summers}}]{VanDam2006}
{van Dam}, M.~A., {Bouchez}, A.~H., {Le Mignant}, D., {et~al.} 2006, \pasp,
  118, 310

\bibitem[{{Van der Swaelmen} {et~al.}(2017){Van der Swaelmen}, {Boffin},
  {Jorissen}, \& {Van Eck}}]{VanderSwaelmen2017}
{Van der Swaelmen}, M., {Boffin}, H.~M.~J., {Jorissen}, A., \& {Van Eck}, S.
  2017, \aap, 597, A68

\bibitem[{{Vogt} {et~al.}(1994){Vogt}, {Allen}, {Bigelow}, {Bresee}, {Brown},
  {Cantrall}, {Conrad}, {Couture}, {Delaney}, {Epps}, {Hilyard}, {Hilyard},
  {Horn}, {Jern}, {Kanto}, {Keane}, {Kibrick}, {Lewis}, {Osborne},
  {Pardeilhan}, {Pfister}, {Ricketts}, {Robinson}, {Stover}, {Tucker}, {Ward},
  \& {Wei}}]{Vogt1994}
{Vogt}, S.~S., {Allen}, S.~L., {Bigelow}, B.~C., {et~al.} 1994, in \procspie,
  Vol. 2198, Instrumentation in Astronomy VIII, ed. D.~L. {Crawford} \& E.~R.
  {Craine}, 362

\bibitem[{{Wang} {et~al.}(2015){Wang}, {Ruffio}, {De Rosa}, {Aguilar}, {Wolff},
  \& {Pueyo}}]{jasonwang2015}
{Wang}, J.~J., {Ruffio}, J.-B., {De Rosa}, R.~J., {et~al.} 2015, {pyKLIP: PSF
  Subtraction for Exoplanets and Disks}, Astrophysics Source Code Library, , ,
  ascl:1506.001

\bibitem[{{Williams} {et~al.}(2009){Williams}, {Bolte}, \&
  {Koester}}]{Williams2009}
{Williams}, K.~A., {Bolte}, M., \& {Koester}, D. 2009, \apj, 693, 355

\bibitem[{{Wizinowich} {et~al.}(2000){Wizinowich}, {Acton}, {Shelton},
  {Stomski}, {Gathright}, {Ho}, {Lupton}, {Tsubota}, {Lai}, {Max}, {Brase},
  {An}, {Avicola}, {Olivier}, {Gavel}, {Macintosh}, {Ghez}, \&
  {Larkin}}]{Wizinowich2000}
{Wizinowich}, P., {Acton}, D.~S., {Shelton}, C., {et~al.} 2000, \pasp, 112, 315

\bibitem[{{Wizinowich} {et~al.}(2006){Wizinowich}, {Le Mignant}, {Bouchez},
  {Campbell}, {Chin}, {Contos}, {van Dam}, {Hartman}, {Johansson}, {Lafon},
  {Lewis}, {Stomski}, {Summers}, {Brown}, {Danforth}, {Max}, \&
  {Pennington}}]{Wizinowich2006}
{Wizinowich}, P.~L., {Le Mignant}, D., {Bouchez}, A.~H., {et~al.} 2006, \pasp,
  118, 297

\bibitem[{{Wright} {et~al.}(2010){Wright}, {Eisenhardt}, {Mainzer}, {Ressler},
  {Cutri}, {Jarrett}, {Kirkpatrick}, {Padgett}, {McMillan}, {Skrutskie},
  {Stanford}, {Cohen}, {Walker}, {Mather}, {Leisawitz}, {Gautier}, {McLean},
  {Benford}, {Lonsdale}, {Blain}, {Mendez}, {Irace}, {Duval}, {Liu}, {Royer},
  {Heinrichsen}, {Howard}, {Shannon}, {Kendall}, {Walsh}, {Larsen}, {Cardon},
  {Schick}, {Schwalm}, {Abid}, {Fabinsky}, {Naes}, \& {Tsai}}]{Wright2010}
{Wright}, E.~L., {Eisenhardt}, P.~R.~M., {Mainzer}, A.~K., {et~al.} 2010, \aj,
  140, 1868

\bibitem[{{Yelda} {et~al.}(2010){Yelda}, {Lu}, {Ghez}, {Clarkson}, {Anderson},
  {Do}, \& {Matthews}}]{Yelda2010}
{Yelda}, S., {Lu}, J.~R., {Ghez}, A.~M., {et~al.} 2010, \apj, 725, 331

\end{thebibliography}
